\documentclass[11pt]{article}
\usepackage{amsmath}
\usepackage{amssymb}
\usepackage{graphicx}
\usepackage{dcolumn}
\usepackage{bm}

\newcommand{\bs}{\boldsymbol}
\newcommand{\la}{\langle}
\newcommand{\ra}{\rangle}

\newcommand{\beq}{\begin{eqnarray}}
\newcommand{\eeq}{\end{eqnarray}}
\renewcommand{\d}{\partial}
\newcommand{\bfl}{\begin{flushleft}}
\newcommand{\efl}{\end{flushleft}}

\usepackage[titletoc]{appendix}
\renewcommand{\title}[1]{%
	\begin{center} \Large \bf #1 \end{center}%
	}
\renewcommand{\author}[2]{%
	\begin{center} {#2}  \vspace{2mm}\\ %
	  \it #1%
	\end{center}%
	\addvspace{\baselineskip}%
	}

\begin{document}

\title{Proper-time Quantum Mechanics for 
Multi-Quark System 
and Composite-Hadron Spectroscopy}

\vspace{.5em}

\author
{%
Research Institute of Science and Technology, College of Science and Technology, \\
Nihon University, Tokyo 101-8308, Japan
}%
{%
Shin Ishida and Masuho Oda
}%

\author
{%
Department of Science and Manufacturing Technology, 
Junior College Funabashi Campus, \\
Nihon University, Funabashi 274-8501, Japan
}%
{%
Tomohito Maeda and Kenji Yamada
}%
\vspace{1em}
\begin{abstract}
One of the most important problems in hadron physics 
is to establish the Lorentz-invariant classification scheme 
of composite hadrons, extending the 
framework of non-relativistic quark model. We present 
an attempt, by developing proper-time $\tau$ quantum mechanics 
on a multi-quark system in particle 
frame (with constant boost velocity $\bs{v}$). 
We start from the 
variational method on a classical mechanics action where 
a constituent quark has Pauli-type $SU(2)_{\sigma}$ spin. 
Then the $SU(2)_{\mathfrak{m}}$ symmetry, 
concerning the sign-reversal on quark mass, 
has arisen with the basic vectors, 
the normal Dirac spinor with $J^{P}=(1/2)^{+}$ and 
the {\it chiral} one with $J^{P}=(1/2)^{-}$, appearing 
as a ``shadow'' of the former. Herewith, the mass reversal 
between these basic vectors become equivalent to the 
chirality, which is a symmetry of the standard gauge 
theory. We describe the role of chirality in hadron 
spectroscopy and regard it as 
attribute \{$\chi$\} of ``elementary'' hadrons in addition to \{$J,P,C$\}. 
A novel feature of our hadron spectroscopy 
is, in the example of $q\bar{q}$ meson system, 
that the ``Regge trajectories'', 
are given by 
{\it mass-squared} vs. the number of quantum $N$; where 
$M^2 =M_{0}^2 +2N\Omega$ 
($N=2n$, $n$ the radial quantum number, $\Omega$ the oscillator quantum),  
and the intrinsic spin of hadrons $\boldsymbol{J}$ comes only 
from quark spin $\boldsymbol{S}$, $\boldsymbol{J}=\boldsymbol{S}$. 
Some phenomenological facts crucial to its validity 
are pointed out on the 
light-through-heavy quarkonium system. 
\end{abstract}
\newpage
\tableofcontents
\newpage
\section{\label{sec0}Introduction 
-- $\tau$-Quantum Mechanics and ``Elementary Hadron''}
The QCD/Standard Gauge-Theory is now regarded to be 
established as the basic under-ground theory of strong interactions among Hadrons. 
However, there is no rigorous technique to treat Bound-State Problem, which leads us to the seriously unsatisfactory situations as follows: 
Firstly for obtaining numerical predictions, only way seems to make computer-calculus, based on Lattice-Gauge Theory, separately in case by case. Secondly its more vital problem is that there could, by no means, draw any concrete and intuitive-picture on such fundamental problems as 
``{\it Quark Confinement}'', 
Picture of  ``{\it Composite-Hadrons}", and 
``{\it Vacuum Condensation}" etc. 
In this work, we shall give some arguments, in replying to all 
the above problems. One of the reason why it may be afforded 
to us is that our method of $\tau$-Quantum Mechanics, 
having the scope of whole possibly boosted-Particle-Frame 
with $\bs{v}$ ($-1<v_i <1$), has naturally led us to a new view-point 
on {\it Compositeness of Hadrons}. In aid of this 
we define a new type of Composite-Hadrons with the definite 
picture as follows: 

\vspace{1em}

({\it Elementary Hadrons})\\
 \quad A Multi-Quark System bounded by Chirality Symmetric 
QCD interaction. \\
 \quad {\it The System by Itself} constructing an {\it Inertial Frame} 
with {\it Non $L$-Definite States}, $L$ being orbital-angular momentum. 

\vspace{1em}

Herewith it is notable that the notion of Elementary Hadron 
is Lorentz-invariant, which originates from its character of 
being an inertial system. And the $\tau$-Wave Function of 
Elementary Hadron becomes tensor-product of the Basic 
vector (Extended Dirac spinor), which plays a role of the 
$\tau$-Wave Function of Constituent-Quark. 
Thus the framework of Composite-Hadron / Elementary-Hadron 
Spectroscopy to be presented in this work should be 
Lorentz-Invariant{\footnote{
See also the arguments in \S \ref{sec5.2.1}. 
}}.

\section{Kinematical Framework and New Physics in the 
Classification Scheme of Hadrons} 
\label{sec1}
Firstly in section \ref{sec1.1}, we give a brief history of our way of 
the extension on the kinematical framework, that is, from non-relativistic scheme, 
$SU(2)_{\sigma}\otimes O(3)_{L}$ (aside from the flavor freedom), to 
covariant scheme\cite{Ref0}, $\widetilde{U}(4)_{DS,\mathfrak{m}}\otimes 
O(2)_{\boldsymbol{r} \perp \boldsymbol{v}}$ 
(the former is a tensor-space of Dirac-spinor embedded with a static 
spin-symmetry $SU(2)_{\mathfrak{m}}$, ${\mathfrak{m}}$ representing 
a new mass-reversal symmetry reflecting the physical situation of 
quark confinement; while the latter is a space of 2-dimensional internal 
spatial-vector $\boldsymbol{r}$, being orthogonal to the boost velocity 
$\boldsymbol{v}$, embedded in the Lorentz-space $O(3,1)_{\rm Lor.}$). 
Secondly in section \ref{sec1.2}, we describe a role of the chirality symmetry in 
composite hadrons, which is valid through light to heavy quark system. 
It is a symmetry of QCD/standard gauge theory, of which importance 
in hadron spectroscopy has been overlooked for many years. 
\subsection{Kinematical Framework of Covariant Classification Scheme for 
Composite Hadrons} 
\label{sec1.1}
\subsubsection{
Most Natural Way of Nonrelativistic-to-Covariant Extension 
and its Difficulty
} 
\label{sec1.1.1}
Most natural way{\footnote{
Strictly, $SU(6)_{\sigma F}\to SU(6)^{(q)}_{\sigma F}\otimes SU(6)^{(\bar{q})}_{\sigma F}$, 
$\widetilde{U}(4)_{DS}\to\widetilde{U}(4)^{(q)}_{DS}\otimes\widetilde{U}(4)^{(\bar{q})}_{DS}$, etc. 
This form of separate symmetry between quark($q$) and anti-quark($\bar{q}$) is certainly 
a requisite result in the relevant spectroscopy based on $\tau$-quantum mechanics, as will be 
seen from Eq.(\ref{215}) in \S \ref{sec2} and in Appendix A.
}}{\cite{Ref4,Ref5,Ref5-2,Ref5-3}} of 
extension is to extend separately both of external and 
internal parts of the kinematical framework is, as 
\begin{subequations}
\beq
\label{Eq1a}
SU(6)_{\sigma F}\otimes O(3)_{L} \ \rightarrow \ \widetilde{U}(12)_{SF}{\mbox{\cite{Ref4}}} \ \otimes \ 
O(3,1)_{\rm Lorentz}{\mbox{\cite{Ref5}}},
\\
\label{Eq1b}
SU(6)_{\sigma F}=SU(2)_{\sigma}\otimes U(3)_{F}, \ \ \widetilde{U}(12)_{SF}=
\widetilde{U}(4)_{DS}\otimes U(3)_{F}, 
\eeq
with the constraint on the expectation value of internal space-time coordinates (momenta), 
$r_{\mu}$($p_{\mu}$) in the definite external momentum $P_{\mu}$-state, as 
\beq
\label{Eq1c} 
\mbox{``Relative-time frozen'' condition:} \ 
\langle P_{\mu}r_{\mu}\rangle=\langle P_{\mu}p_{\mu}\rangle =0. 
\eeq
\end{subequations}
However, this way had been seen to be closed by two 
No-Go theorems{\cite{Ref6,Ref7}}, as 
\begin{subequations}
\beq
\label{Eq2a}
\mbox{External:~}SU(6)_{\sigma F} &\nrightarrow&   \ \ 
\mbox{Relativistic Extension \ \ by Coleman-Mandula{\cite{Ref6}}}, \\
\label{Eq2b}
\mbox{Internal:~} O(3)_{L} &\nrightarrow&   \  \ O(3,1)\ \ 
\mbox{Yukawa's Bi-local Field Theory{\cite{Ref7}}}, 
\eeq
\end{subequations}
where the origin of internal No-Go theorem is considered to come from the 
close connection between the {\it external} coordinate $X_{\mu}$ and 
the {\it internal} coordinate $r_{\mu}$, 
as is seen from an ad hoc subsidiary ``relative-time frozen'' condition. 
\subsubsection{
Semi-Phenomenological Derivation of Covariant Classification Scheme 
of Composite Hadrons
} 
\label{sec1.1.2}

Then we have chosen a semi-phenomenological means 
of pass-through{\footnote{
Exciton picture of quarks was first proposed by 
O. Hara and T. Goto\cite{Hara-Goto}. 
Successively, on the basis of this picture, 
the present author\cite{Ref5-2} 
had presented the kinematical 
framework for systematic hadron phenomenology 
which is an origin of line of thoughts in this work. 
Actually the Covariant Oscillator Quark Model (COQM){\cite{IshidaYamada1987}} 
with this framework had been applied for many years to various 
phenomena of hadron physics. 
}}. Concerning the intrinsic quark-spin, 
{\it a new freedom of $SU(2)$-mass spin} is supposed{\footnote{
Embedding of non-covariant $SU(2)_{\mathfrak{m}}$-spin into the 
covariant $\widetilde{U}(4)_{DS,\mathfrak{m}}$-space had been done, 
as a static-unitary symmetry, in Ref.{\cite{Ishida-Roman}}.
The existence of $SU(2)_{\mathfrak{m}}$-symmetry is deduced from 
the Klein-Gordon equation as master equation, to be satisfied by 
constituent-quark field(, see Eq.(\ref{Eq9})).}} 
in addition to the 
$SU(2)$-$\sigma$ spin, which makes possible to apply the crossing rule 
to confined quarks, in conformity with {\it color-singlet condition of parent 
hadrons}; while concerning the undesirable connection of internal $r_{\mu}$ 
to external $X_{\mu}$, 
it is separated{\footnote{This implies that the internal 
space-time variable $r_{\mu}$ 
concerns only mass term of the Yukawa-type Klein-Gordon equation for relevant composite 
hadrons, as will be described in section \ref{2.2}.
}} by supposing $r_{\mu}$ to be spacelike and 
$\boldsymbol{r}$, also, to be {\it orthogonal} 
to the boost velocity $\boldsymbol{v}$. 

The corresponding formulas and remarks to the above are systematically 
given as: \\
$[$On external freedom$]$ the covariant spin-classification scheme, 
$\widetilde{U}(4)_{DS}$, is extended to $\widetilde{U}(4)_{DS,\mathfrak{m}}$, 
introducing a new static-unitary symmetry\cite{Ishida-Roman} $SU(2)_{\mathfrak{m}}$, 
embedded at Observer Frame ($\boldsymbol{v}=\boldsymbol{0}$), as\cite{Ishida2000, Ishida2002} 
\begin{subequations}
\label{Eq3}
\beq
\label{Eq3a}
\mbox{Basic Urciton Spinors} &:& \{U_{r+}(P,M_{+})({\frac{1}{2}}^{+}),
U_{r-}(P,M_{-})({\frac{1}{2}}^{-}) \}, \\
\label{Eq3b}
\mbox{Basic anti-Urciton Spinors } &:& \{\bar{V}_{\bar{r}+}(P,M_{+})({\frac{1}{2}}^{-}),
\bar{V}_{\bar{r}-}(P,M_{-})({\frac{1}{2}}^{+}) \},
\eeq
\end{subequations}
\beq
\label{Eq4}
U_{r-}\equiv -\gamma_{5}U_{r+}, \ \bar{V}_{\bar{r}-}\equiv \bar{V}_{\bar{r}+}\gamma_{5},
\eeq
where $r$($\bar{r}$)$=\left\{+,-\right\}$ is the 
direction of $\hat{m}_{3}$($\hat{\bar{m}}_{3}$) for urciton 
(anti-urciton) to be called $\left\{\right.$Paulon, Chiralon$\left.\right\}$, respectively. 
Here it is noted that the 
freedom of Chiralon predicts the existence of {\it light-mass}( -- in the non-relativistic 
scheme) particles ($\sigma (500,0^{+})$, 
$\rho/\omega (1250,1^{-})$, $N(1440,{1/2}^{+})$, $\Lambda(1405,{1/2}^{-})$). 

[On internal freedom] 
the undesirable connection between the variables 
$r_{\mu}$ and $X_{\mu}$ is separated, so that our interests are only in the 
space-like region of $r_{\mu}$, and then supposing its spatial component $\boldsymbol{r}$ 
be orthogonal to boost-velocity $\boldsymbol{v}$ as 
\begin{subequations}
\beq
\label{Eq5a}
P_{\mu}r_{\mu}=0, \ \ {\mbox{and}}  \ \ \boldsymbol{v}\cdot \boldsymbol{r}=0. 
\eeq
Thus our relevant internal space becomes as 
\beq
\label{Eq5b}
O(3,1)_{\rm Lor.} \ \Rightarrow \ O(2)_{\boldsymbol{r}\perp\boldsymbol{v}} \ 
\mbox{embedded in}  \ O(3,1)_{\rm Lor}. 
\eeq
\end{subequations}
Here, it is also noted that there exists no relative-time problem. 

Accordingly, the covariant framework of our hadron-classification scheme (aside from 
the flavor freedom) becomes,{\footnote{
In Eq.(\ref{Eq6}) is written the dimension two of {\it general internal extension}, 
as $O(2)_{\bs{r}\perp \bs{v}}$. However, in the {\it actual problem} will be adopted 
the one-dimensional oscillator, might be noted as $O(1)_{\bs{r}\perp \bs{v}}$(, see 
around Eqs.(\ref{70d}) and (\ref{Hooke})). 
}} as 
\beq
\label{Eq6}
\{\widetilde{U}(4)^{(q)}_{DS,\mathfrak{m}}\otimes\widetilde{U}(4)^{(\bar{q})}_{DS,\mathfrak{m}}\}\otimes O(2)_{\boldsymbol{r}\perp\boldsymbol{v}} {\rm \ in} \ O(3,1)_{\rm Lor} . 
\eeq
\subsection{Chirality -- the Symmetry in QCD Gauge Theory and its Role 
in Hadron-Spectroscopy} 
\label{sec1.2}
The chirality is an important symmetry 
conserved through all types of the QCD gauge interaction/{\it upper-ground} 
strong interaction with each flavored quarks. 
This property of chirality seems us that it deserves to be an attribute 
of, somewhat {\it elementary entity, } Composite Hadrons, consisting of 
{\it confined quarks}. 

They shall be called as {\it Elementary Hadrons} in this work. 

\subsubsection{Chirality Conservation in QCD
} \label{sec1.2.1}

The effective interaction Lagrangian ${\mathcal L}_{I} (q(x)_{i}\mbox{'s})$ for 
any kind of vertices among quark-pairs, $(\bar{q},q)_{i}$'s, are obtained by substituting, 
in the relevant free-quark Lagrangians, the covariant derivative $D_{\mu}\equiv
\partial_{\mu}-i g A_{\mu}$($A_{\mu}\equiv \sum_{a=1}^{8}A_{\mu}^{a}
\lambda^{a}/2$ in QCD) for 
the normal derivative $\partial_{\mu}$. Accordingly, the chirality conservation is 
easily seen, as follows: \\
\mbox{{Substitution}}
\begin{subequations}\\
\beq
\label{Eq7a}
\bar{q}_{i}\gamma_{\mu}\partial_{\mu}q_{i} 
\rightarrow \bar{q}_{i}\gamma_{\mu}D_{\mu}q_{i}=
\bar{q}_{i}\gamma_{\mu}\partial_{\mu}q_{i}-ig \bar{q}_{i}\gamma_{\mu}A_{\mu}q_{i}, 
\eeq
\beq
\label{Eq7b}
\Rightarrow \mathcal{L}_{I}(q_{i}\mbox{'s})=
\mathcal{L}_{I}((\bar{q}\gamma_{\mu}A_{\mu}q)_{i}\mbox{'s}).
\eeq
\end{subequations}
\mbox{{Chirality Transformation}}
\begin{subequations}
\beq
\label{Eq8a}
q_{i}\rightarrow q_{i}^{\chi}\equiv \hat{\chi}q_{i}=-\gamma_{5}q_{i},  \\
({\rm For \ Space \mathchar`-Time \ Reflection }\ R_{PT}:X^{'}_{\mu}=-X_{\mu})\nonumber
\eeq
\beq
\label{Eq8b}
\mathcal{L}_{I}(q_{i}^{\chi}\mbox{'s})=\mathcal{L}_{I} (q_{i}\mbox{'s}). 
\eeq
\end{subequations}
\subsubsection{Role of Chirality in Elementary Hadrons and 
Mass Reversal of Constituent Quarks}
\label{sec1.2.2}

The contents of this subsection is an introduction of 
the kinematical framework, 
{\it Evolved COQM}~(Covariant Oscillator Quark Model) with 
chirality-symmetric constituent quarks, to be developed in the 
next section \S \ref{sec2}. Here, it is to be noted that the old 
COQM\cite{IshidaYamada1987} had not yet the notion of {\it chirality}.
\paragraph{
({\it The case of constituent-quark field})
}
\begin{subequations}
\beq
\label{Eq9a}
&&\mbox{{Wave Equation:}}\hspace{29em} \nonumber \\
&& \ \ \ \ \left(\frac{\partial^2}{\partial X_{\mu}^2}-M^2\right) \Phi_{\alpha}(X)=0, \\
\label{Eq9b}
&& \ \ \ \ \left(\gamma_{\mu}\frac{\partial}{\partial X_{\mu}}+M_{\pm}\right) \Phi_{\alpha}(X)=0 \ (M_{\pm}\equiv\pm M), 
\eeq
\label{Eq9}
\end{subequations}
\begin{subequations}
\label{Eq10}
\beq
\label{Eq10a}
&&\mbox{{Wave Function:}} \nonumber \hspace{29em}\\
&& \ \ \ \ \Phi_{\alpha}(X)=\sum_{P_{\mu}(P_{0}=E)}\left(\sum_{r=\pm}U_{r,\alpha}(P,M)e^{iPX}
+\sum_{\bar{r}=\pm}V_{\bar{r},\alpha}(P,M)e^{-iPX}
\right)\\
&& \ \ \ \ \hspace{3.2em} \equiv \sum_{P_{\mu}(P_{0}=E)}\left(W_{\alpha}(P,M)^{(q)}e^{iPX}
+W_{\alpha}(P,M)^{(\bar{q})}e^{-iPX}
\right), 
\label{Eq10b}
\eeq
\end{subequations}
where the intrinsic spin-wave functions in $\widetilde{U}(4)_{\mathfrak{m}}$-spin scheme, 
$W_{\alpha}^{(q)}$ and $W_{\alpha}^{(\bar{q})}$, of constituent 
quarks and 
anti-quarks, respectively, are defined. 
The Klein-Gordon Eq.(\ref{Eq9a}) has been set up as the master wave equation. The 
reason is that the {\it observable entity} is not quarks but parent hadrons, satisfying it 
{\it as the mass-shell} condition. It is interesting that this equation leads to Eq.(\ref{Eq9b}), 
which gives the two basic-vectors of the $SU(2)_{\mathfrak{m}}$-symmetry space, Paulons and 
Chiralons. Herewith, the wave function of constituent quark field $\Phi_{\alpha} (X)$ 
is expanded{\footnote{Here it may be notable that the sign of mass, which 
had been meaningless 
for free Dirac particles, now plays an important role for {\it confined quarks}; 
and that {\it inertial frame}(, with respect to Lorentz Transformation) with 
definite boost velocity ($\bs{v}\neq \bs{0} $) 
for isolated, multi-particle system seems to be well representing the physical 
situation of quark confinement. The frame with $\bs{v}\neq \bs{0} $ 
is called the particle frame, while the one with $\bs{v}= \bs{0} $ 
the observer frame. The formulas obtained in the former (latter) 
becomes Lorentz-invariant (-covariant). 
}} as in Eq.(\ref{Eq10}), where the sum on $r$($\bar{r}$) (direction of $\hat{\mathfrak{m}}_{3}$, 
$\hat{\bar{\mathfrak{m}}}_{3}$) guarantees the {\it mass-reversal symmetry} of 
$\Phi_{\alpha}(X)$. 
In relation with this symmetry it should be noted that the $\Phi_{\alpha}(X)$ 
plays a role of basic asymptotic state of $S$-matrix for strong interaction 
among general {\it elementary hadrons in the 
$\widetilde{U}(4)_{DS,\mathfrak{m}}$-spin 
multiplets}(, as will be shown in \S \ref{sec-2015-5-26}).

The Dirac-spinors given in Eq.(\ref{Eq3}) and (\ref{Eq4})  play the two side-roles in the 
$\widetilde{U}(4)_{DS,\mathfrak{m}}$-classification scheme, 
satisfying, respectively, the following equations: \\
As Constituent Quark spinors, 
\begin{subequations}
\beq
\label{Eq11a}
(iP_{\mu}\gamma_{\mu}\pm M)U_{r\pm}(P,M)=0, \ 
(iP_{\mu}\gamma_{\mu}\mp M)V_{\bar{r}\pm}(P,M)=0, 
\eeq
\mbox{and as Urciton spinors, } 
\beq
\label{Eq11b}
\ (i u_{\mu}\gamma_{\mu}\pm 1)U_{r\pm}(\boldsymbol{v}(P))=0, \ 
(i u_{\mu}\gamma_{\mu}\mp 1)V_{\bar{r}\pm}(\boldsymbol{v}(P))=0, 
\eeq
\end{subequations}
where these equations (\ref{Eq11a}, \ref{Eq11b}) are connected simply by 
$(P_{\mu}=Mu_{\mu}; \boldsymbol{v}\equiv \frac{d \boldsymbol{X}}{d T}, u _{\mu}\equiv 
\frac{d X_{\mu}}{d \tau})$. The constituent quark-spinor in Eq.(\ref{Eq11a}) concerns the {\it mass-reversal structure} 
of parent hadrons, while the urciton-spinor in Eq.(\ref{Eq11b}) does their {\it chirality structure}, as will be shown in the case of $q\bar{q}$-meson(, see Eqs.(\ref{Eq17}) 
and (\ref{Eq18})). 
Here it should be noted that, for the wave function of constituent quark-field 
$\Phi (X)$ in Eq.(\ref{Eq10}), 
the two transformations of mass reversal and of chirality become equivalent, 
as is easily seen from 
Eq.(\ref{Eq11a}) and their definition: \\
\begin{subequations}
{\it Equivalence of mass reversal} and {\it chirality}
\beq
\label{Eq12a}
\mbox{Mass Reversal{\cite{Ref9}}~~~~~} \hat{R}_{m}=\rho_{1}(\mathfrak{m}): M_{\pm}\to M_{\mp}, \ 
\Phi_{\pm}(X)\to \Phi_{\mp}(X), \\
\label{Eq12b}
\mbox{Chirality Transformation{\cite{Ref8}}~~~~} \hat{R}_{\rm PT}=-\gamma_{5}: 
X_{\mu}\to -X_{\mu}, \ \Phi (X_{\mu})\to \Phi (-X_{\mu}).
\eeq
\end{subequations}
It may be also instructive to point out that there is two types of representation for the 
urciton-spinors, as: \\
\begin{subequations}
{\it Two representations of urciton-spinor}
\beq
\label{Eq13a}
U_{r\pm, \alpha}: && \mbox{Bargmann-Wigner representation} \ \ r\pm=\mbox{eigen-value of } 
{\rho}_{3} (\mathfrak{m}), \\
\label{Eq13b}
U_{\alpha}^{(\chi\pm)}: && \mbox{Chirality representation}  
\ \ \chi \pm=\mbox{eigen-value of }
{\chi}.
\eeq
\end{subequations}
They are related with each other by 
\beq
\label{Eq14}
U_{\alpha}^{(\chi\pm)}=\frac{1}{\sqrt{2}}(U_{r+}\pm U_{r-})_{\alpha}, \  \ 
U_{r\pm, \alpha}=\frac{1}{\sqrt{2}}(U^{(\chi +)}\pm U^{(\chi -)})_{\alpha}. 
\eeq
\paragraph{
({\it The case of $(q\bar{q})$-meson field}) 
}
{\it Wave Equation and wave function} \\
The Klein-Gordon wave equation has been set up so as to
represent the mass-shell condition of mesons as 
\beq
\label{Eq15}
\left(\frac{\partial^2}{\partial X_{\mu}^2}-M^2\right)\Phi_{\alpha}^{(M)\beta}(X)=0.  
\eeq
The meson wave function 
$\Phi_{\alpha}^{(M)\beta}(X)$ is expanded as 
\beq
\Phi_{\alpha}^{(M)\beta}(X)&=&
\sum_{P_{\mu}(P_{0}=E)}\left[\sum_{r,\bar{r}}U_{\alpha, r}(\boldsymbol{v}(P))\bar{V}_{\bar{r}}^{\beta}(\boldsymbol{v}(P))
e^{iPX}+\sum_{r,\bar{r}}V_{\alpha,\bar{r}}(\boldsymbol{v}(P))\bar{U}_{{r}}^{\beta}
(\boldsymbol{v}(P))e^{-iPX}\right] \nonumber \\ 
&\equiv &
\sum_{P_{\mu}(P_{0}=E)}\left[W_{\alpha}^{(M)\beta}(\boldsymbol{v}(P))
e^{iPX}+W_{\alpha}^{(\bar{M})\beta}(\boldsymbol{v}(P))e^{-iPX}\right],  
\label{Eq16}
\eeq
where the intrinsic-spin wave functions 
$W_{\alpha}^{(M)\beta}(\boldsymbol{v}(P))$ and 
$W_{\alpha}^{(\bar{M})\beta}(\boldsymbol{v}(P))$ in the $\widetilde{U}(4)_{\mathfrak{m}}$-spin scheme 
are also defined. Here it is to be noted that summation on the 
suffices $r$($\bar{r}$) guarantees mass-reversal symmetry, and/or chirality symmetry of 
the relevant meson-WF. \\
{\it Expansion of meson WF into the definite-$J^{PC\chi_{m}}$-members{\footnote{
\label{foot8}
From here; in order to discriminate {\it the relevant chirality, originating in 
$\hat{R}_{m}$}, from {\it that, operating on pure indices of higher-spin WF}; 
the label $\chi_{m}$ will be used for the former (As for details, see \S \ref{sec:4-4}).
}}} \\
The general meson-WF is expanded into complete set of the members with 
definite chirality $\{\chi_m \}$, new attribute of elementary composite hadrons in addition to 
$\{J^{PC} \}$, 
\beq
\label{Eq17}
\Phi_{A}^{(M)B}(X)=\sum_{i}\phi_{a}^{(i)b}(X)\Gamma_{\alpha}^{(i)\beta},
\eeq
where $A\equiv \{\alpha, a\}$ etc. ($a$ denoting flavor index) and the appeared quantities are defined as\footnote{The $\tilde{\gamma}_{\mu}$ is 
defined so as to be $u_{\mu}\tilde{\gamma}_{\mu}=0.$} 
\begin{subequations}
\beq
\phi^{(i)}&=&\{ P_{s}^{(N)}, P_{s}^{(E)}, S^{(N)}, S^{(E)}, 
V_{\mu}^{(N)}, V_{\mu}^{(E)}, A_{\mu}^{(N)}, A_{\mu}^{(E)}\}, \\
\Gamma^{(i)}&=&\{ i\gamma_{5}, i\gamma_{5}(i u_{\nu}\gamma_{\nu}), 1, iu_{\nu}\gamma_{\nu}, 
i\tilde{\gamma}_{\mu}, -i{\sigma}_{\mu\nu}u_{\nu}, 
i\gamma_{5}\tilde{\gamma}_{\mu},  
-i\gamma_{5}{\sigma}_{\mu\nu}u_{\nu} \}, \\
\chi_{m}^{(i)}&=&\{-,+,-,+,+,-,+,- \}. 
\label{Eq18c}
\eeq
\label{Eq18}
\end{subequations}
The Eq.(\ref{Eq17}) is the covariant quark-representation of composite $q\bar{q}$-meson system, 
proposed in Ref.\cite{Ishida2000}, which started the relevant covariant classification scheme. 
Here $\phi^{(i)}(X)$'s with the definite Lorentz transformation property in addition to the attributes 
$\{J^{PC\chi_{m}}\}$ are to be second-quantized{\footnote{
It is to be noted that Urciton spinors in BW representation Eq.(\ref{Eq13a}) appear as the 
spin WF of mixed-states of the quantized pure-states with definite $\chi$, 
(See, Eq. (\ref{Eq14}) ). 
}}. (See, \S \ref{Quantum Mechanics}.)
\section{Proper-time Quantum Mechanics for Multi-Quark System and Quantization 
of Composite-Hadron Fields} 
\label{sec2}
In section \ref{sec2}, proper-time $\tau$-quantum mechanics for 
multi-body confined-quark system and quantization of Composite 
Hadron field is developed. The similar to conventional procedures 
are performed, but all in a certain 
Inertial Frame (with $\boldsymbol{v}=\rm const$); starting from an 
application of variational method to a classical action of the 
relevant confined system, where quarks have Pauli-type 
$SU(2)_{\sigma}$-intrinsic spin. 
A notable feature of the $\tau$-Quantum Mechanics is, it is concerned only 
future-development : which induces application of the 
crossing rule for ``Negative-Energy Problem''. 
This rule is conventionally supposed ad hoc. 
The $\tau$-Quantum Mechanics also induces Existence of the Chiral-quark, 
with $J^P=(1/2)^{-}$ (, as a ``Shadow'' of 
Pauli-quark with $J^P=(1/2)^{+}$), which 
is considered to be an origin of new 
exotics, mysterious from non-relativistic scheme. 
These parity-doublet pair of quarks with $J^{P}={\frac{1}{2}}^{(\pm)}$ play rightly 
the role of Basic Vectors of the covariant $\widetilde{U}(4)_{DS,\mathfrak{m}}$ 
spinor-space, semi-phenomenologically derived in \S \ref{sec1.2.1}. 
The Non-Local $\tau$-Wave Function for relevant multi-quark system 
leads, in Observer F., to a complete set of the corresponding Local Hadron-Fields 
in the $\widetilde{U}(4)_{DS,\mathfrak{m}}$-spinor scheme. 
The title, ``Prototype Mechanics for Solitary Urciton-Quark Field/
Prototype Mechanics for Multi-Urciton Quark Fields{
\footnote{Prototype Mechanics for the Bi-local Field Theory had been introduced 
by T. Takabayasi\cite{Takabayasi}. See also, ``Mechanical Model of 
Bi-local Field Theory'' by T. Goto\cite{Goto}. }
}'', for the subsections \S \ref{2.1} and \S \ref{2.2}, 
reflects this situation. 

\begin{figure}[htbp]
\centering
  \mbox{
   \includegraphics[width=15cm]{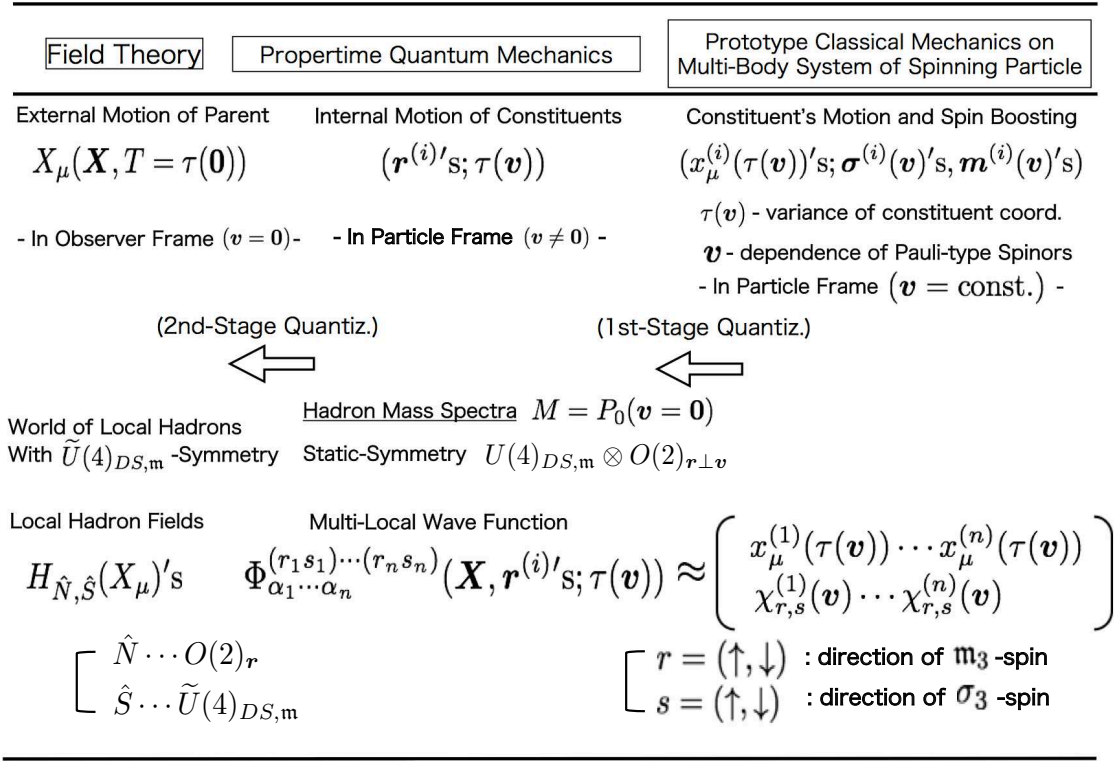}
   }
   \caption{Overview for Quantum Field Theory of Composite Particles. 
The two prescriptions playing an important role, 
i) Two-stage(1st and 2nd) Quantization of wave function at the proper time 
$\tau (\bs{v}\neq \bs{0})$ and $\tau (\bs{0})=T$, respectively, 
for a single, confined multi-quark system. 
ii) Special Definition of the internal coordinates $r_{\mu}^{(i)}\equiv 
x_{\mu}^{(i)}-X_{\mu}$, measured from the external coordinate $X_{\mu}\{ \bs{X}
\equiv \sum_{i}m^{(i)}\bs{x}^{(i)}/M, \tau (\bs{v})\}$. 
 The freedom of mass-spin has arisen, reflecting the situation of 
Quark-Confinement.  As for details, see the \S \ref{new3-2-1}, 
and Eqs. (\ref{2-9-72a}, \ref{2-9-72b}) in \S \ref{sec3.2}. }
   \label{fig1}
 \end{figure}

Before going into details, is shown in Fig.{\ref{fig1}}; 
Overview on constructing 
Quantum Field Theory of Composite Particles. 
Here, for simplicity, it is concerned the multi-particle system, 
consisting of only quarks, but no anti-quarks. 

In this figure a special prescription on separation between the external 
and internal coordinates of the constituent particles is applied, of 
which Lorentz invariance is guaranteed; as mentioned before, by the relevant 
{\it Particle Frame}(PF) (with non-zero boost-velocity $\boldsymbol{v}\neq 0$)
 itself, being as an inertial frame. (See, the argument on 
``Elementary Hadron'' in \S \ref{sec0}.)

Herewith, especially notable is that the seemingly three-dimensional 
definition of the spatial component $\bs{X}$ is identical to that of 
center of mass coordinates in classical mechanics. Accordingly the external coordinate 
$X_{\mu}$ may be also called as ``center of mass coordinate'' of the multi quark system.  

\subsection{Prototype Mechanics for Solitary Urciton-Quark Field}
\label{2.1}
In the covariant classification scheme, or the precedent COQM,  
general composite hadrons are described{\cite{IshidaYamada1987}} 
by the relevant multi-local fields, 
which are supposed, rather in an ad-hoc way, to satisfy the Yukawa-type 
Klein-Gordon wave equation. Some logical bases for this 
description are provided from the following classical mechanics
{\footnote{
Along this line of thoughts we had made some consideration{\cite{Ishida-Sonoda}} 
on its implication to 
COQM{\cite{IshidaYamada1987}}. 
The contents of this sub-section are results of the further study on this problem. 
}} of a system of confined particles; which leads, through the 1st-stage 
quantization, to the relevant 
Klein-Gordon equation.  

First we shall show its essential points in the 
ideal case of local Klein-Gordon field, 
describing the basic urciton-quark with a definite 
mass in the $\widetilde{U}(4)_{DS, \mathfrak{m}}$ space. 

\subsubsection{Classical Mechanics}  


In classical mechanics the relevant action 
is given by{
\footnote{
It is evident from Eq.~(\ref{2.27}) that sole existence of ``chiral particle'' with the mass, 
$M=-|M|<0$, is prohibited by the {\it least}-action principle. 
}} 
\beq
 S&=&\int^{\tau}_{0}{\mathcal{L}}(\dot{X}_{\mu})d\tau, \ \ {\mathcal{L}}=-M\sqrt{-\dot{X}_{\mu}^2}  \ \ (M>0),
\label{2.27}
\eeq
where $\dot{X}_{\mu}\equiv d X_{\mu}/d\tau$ ($\dot{X}_{\mu}^2=-1$), $d\tau=\sqrt{-(dX_{\mu})^2}=
dT\sqrt{1-\boldsymbol{v}^2}$($\boldsymbol{v} \equiv d \boldsymbol{X}/d T$; boost velocity) and 
$\tau=\int d \tau$ is proper-time defined along world line with the restriction $dX_{0}/d\tau >0$. 
The variation of $S$ is given as
\beq
\delta S= 
\left.\frac{M\dot{X}_{\mu}}{\sqrt{-\dot{X}_{\mu}^2}} \delta X_{\mu}
\right|_{0}^{\tau}-M\int_{0}^{\tau}\left(\frac{d}{d\tau}\frac{\dot{X}_{\mu}}{\sqrt{-\dot{X}_{\mu}^2}} \right)
\delta {X}_{\mu} d\tau. 
\eeq

This leads to Lagrange Equation of Motion and definition of 4-momentum $P_{\mu}$, respectively, as 
\begin{subequations}
\beq
&& \ \ \ \ \ \ \ \ \ \ \ \ \  \frac{d}{d \tau}M \dot{X}_{\mu}=0, 
\ \ \ \ \ \ \ \ \ \ \ \ \  \ \ \ \ \ \ \ \ \ \ \ \ \ 
\eeq
\beq
&& \ \ \ \ \ \ 
\left.P_{\mu}=\frac{\partial S}{\partial X_{\mu}(\tau)}\right|_{\rm Phys.Pass}
=M\dot{X}_{\mu}=M u_{\mu}, \ \ \
\label{four-mom}
\eeq
\end{subequations}
where $u_{\mu}=d X_{\mu}/d\tau$ is the 4-velocity, and $u_{\mu}^2=-1$. 

The above equations lead to conservation of the four-momentum 
\begin{subequations}
\beq
\frac{d}{d \tau}P_{\mu}=0, 
\label{2.28a}
\eeq
where
\beq
\label{2.28b}
P_{i}=M\frac{d X_{i}}{d\tau}=\frac{Mv_{i}}{\sqrt{1-\boldsymbol{v}^2}}, 
\ P_{0}=\sqrt{\boldsymbol{P}^2+M^2}
=M\frac{d X_{0}}{d\tau}=\frac{M}{\sqrt{1-{\boldsymbol{v}}^2}}. \ 
\eeq
\label{2.28}
\end{subequations}
The equation (\ref{2.28b}) is a famous Einstein formula. 
Here it may be instructive to remark that 
the equations (\ref{2.28a}) and (\ref{2.28b}) in the, so to speak, 
PF with non-zero boost-velocity $\boldsymbol{v}\neq 0$ become, respectively, 
the equations in the {\it Observer Frame} (OF) with zero boost-velocity $\boldsymbol{v}= 0$ 
\begin{subequations}
\beq
\frac{d}{d T}M=0, 
\eeq
and 
\beq
\ P_{i}=0, \ P_{0}=M.
\eeq
\end{subequations}

\paragraph{
(\it $\tau$-Hamiltonian and $\tau$-Gauge Condition)
}

In order to take a further step into 
the 1st-stage quantization the relevant physical quantities are represented by canonical 
variables, presently $X_{\mu}$ and $P_{\mu}$.

Firstly Hamiltonian is directly write down{
\footnote{
Eq.~(\ref{2.31a}) is given in the ``general'' gauge with $\tau^{'}$ in Eq.~(\ref{2.32}).
}}, while it is constrained by the {\it gauge condition} $G$; as 
\begin{subequations}
\beq
{\mathcal H}=P_{\mu}\dot{X}_{\mu}-{\mathcal{L}}=\frac{K}{M}(P_{\mu}^2+M^2)=0, 
\label{2.31a}
\eeq
\beq
G\equiv P_{\mu}^2+M^2 \propto {\mathcal H}=0. 
\label{2.31b}
\eeq
\end{subequations}
``{\it Zero}''-${\mathcal H}$ comes from the invariance of the 
action (\ref{2.27}) under a gauge transformation of $\tau$ 
\beq
\tau \to \tau^{'}=K\tau \ \ \ (K>0).
\label{2.32}
\eeq 
This invariance reflects a physical situation that the {\it time scale of an 
isolated confined-quark system is unobservable}. The indefinite factor $K$ is 
determined by requiring that 
the equation of motion be equivalent in both Lagrange's and Hamilton's ones. 
Then the latter equation leads to the solution as 
\beq
\frac{d}{d \tau}X_{\mu}=\frac{\partial {\mathcal H}}{\partial P_{\mu}}=K\frac{2}{M}P_{\mu}, \ \ \
\frac{d}{d \tau}P_{\mu}=-\frac{\partial {\mathcal H}}{\partial X_{\mu}}=0.
\label{2.33}
\eeq
Comparing these, respectively, with Eqs.~(\ref{2.28b}) and (\ref{2.28a}), the {\it $K$-factor} is determined 
as $K=1/2$ and the Hamiltonian $\mathcal H$ is fixed as 
\beq
{\mathcal H}=\frac{1}{2M}(P_{\mu}^2+M^2).
\label{2.34}
\eeq

\subsubsection{Quantum Mechanics}
\label{QMsolo1}
Then our relevant, Lorentz-covariant 1st-quantization may be accomplished by replacing 
Poisson Bracket with commutator between the canonical variables as 
\beq
\{X_{\mu}, P_{\nu}\}_{\rm P.B.}=\delta_{\mu\nu} \to 
\left[\hat{X}_{\mu}, \hat{P}_{\nu}\right]_{\rm com.}=i \delta_{\mu\nu}. 
\label{2.35}
\eeq

\begin{flushleft} 
(\it $\tau$-Schr{\"o}dinger Eq. and $\tau$-Gauge Constraint)
\end{flushleft}

Then the ``Proper-time'' Schr{\"o}dinger Wave Equation will be set generally in 
PF as 
\beq
i\frac{d}{d\tau} \Phi_{\alpha}(\boldsymbol{X}:\tau)&=&
\hat{\mathcal H}(\hat{\boldsymbol{P}})\Phi_{\alpha} (\boldsymbol{X}:\tau), \nonumber\\
\hat{\mathcal{H}}&\equiv &\frac{1}{2M}\hat{G}=\frac{1}{2M}(\hat{P}_{\mu}^2+\hat{\mathcal{M}}^2), 
\label{2.36}
\eeq
while the gauge constraint Eq.~(\ref{2.31b}) is represented in 
OF as{\footnote{{\label{foot14}}
In Eq.~(\ref{2.37}) and accordingly in Eq.~(\ref{2.36}), we have simply 
put a {\it{Dirac-spinor index} $\alpha$} on the relevant WF as $\Phi_{\alpha}$. 
This prescription is guaranteed by the 
fact that the relevant variational method on the internal space $O(2)_{\bs{r}\perp\bs{v}}$ is 
common in all inertial frame with respective boosting velocity $\bs{v}$'s. 
}} 
\beq
\hat{G}\Phi_{\alpha}(X_{\mu}) 
=\left(\hat{P}_{\mu}^2+M^2\right) \Phi_{\alpha}(X_{\mu})=0.
\label{2.37}
\eeq
As will be shown in the next item, the solution of these two equations 
become equivalent to the space-time WF 
of two basic vectors in the $\widetilde{U}(4)_{DS,\mathfrak{m}}$-spinor space, 
given in Eqs.(\ref{Eq3a}) and (\ref{Eq3b}).
This implies that the above equations (\ref{2.36}) and (\ref{2.37}) lead 
rightly to the local Klein-Gordon wave equation Eq.(\ref{Eq9a}), 
which is the master equation to be satisfied by the basic urciton-vectors in the 
$SU(2)_{\mathfrak{m}}$-space. 

\begin{flushleft} 
(\it Concrete Form of Urciton-Quark Wave Function)
\end{flushleft}

For perception of the basic structure of our 
Lorentz-invariant proper time quantum mechanics we show 
a detailed form of the solution, {\footnote{\label{footnote26}
Here it may be instructive to note the two 
implications of the gauge constraint (\ref{2.37}): as \ 
i) $\hat{G}=(-\hat{P}_{0}+\hat{E})(\hat{P}_{0}+\hat{E})$; $\hat{E}\equiv
\sqrt{M^2+\hat{\boldsymbol{P}}^2}$, $i{d \Phi^{(\pm)}}/{d\tau}=\pm \hat{E} \Phi^{(\pm)}$, 
that is, leading to $\Phi^{(+)}$/$\Phi^{(-)}$: positive/negative-$\tau$ frequency 
solutions.  \ 
ii) $\hat{G}=[(i\hat{P}_{\mu}\gamma_{\mu}+M)(i\hat{P}_{\mu}\gamma_{\mu}-M)]$, 
$\Phi_{\alpha}(X,\boldsymbol{v})=\{\Phi^{r+}_{\alpha},\Phi^{r-}_{\alpha}\}$, 
that is , leading to the basic urciton spinors of $SU(2)_{\mathfrak{m}}$-spin space.
}}
with definite momenta $\boldsymbol{P}(\boldsymbol{v})$, 
of both Eq.~(\ref{2.36}) and Eq.~(\ref{2.37}) as
\beq
&&\Phi_{\boldsymbol{P}, \alpha}(\boldsymbol{X};\tau (\boldsymbol{v}))=
\Phi^{(+)}_{\boldsymbol{P}, \alpha}(\boldsymbol{X};\tau (\boldsymbol{v}))+
\Phi^{(-)}_{\boldsymbol{P}, \alpha}(\boldsymbol{X};\tau (\boldsymbol{v})) \nonumber \\
&=&\sum_{r,s}U_{r,s;\alpha}(\boldsymbol{v}(P))
e^{+i\boldsymbol{P}(\boldsymbol{v})\cdot\boldsymbol{X}}e^{-iP_{0}(\boldsymbol{v})\tau (\boldsymbol{v})}
+\sum_{\bar{r},\bar{s}}V_{\bar{r},\bar{s};\alpha}(\boldsymbol{v}(P))
e^{-i\boldsymbol{P}(\boldsymbol{v})
\cdot\boldsymbol{X}}e^{+iP_{0}(\boldsymbol{v})\tau (\boldsymbol{v})}, \nonumber\\
\label{2.38}
\eeq
where the respective terms in the second and the last equality corresponds 
to each other, and 
$P_{\mu}(\boldsymbol{P}(\boldsymbol{v}),P_{0}(\boldsymbol{v}))$ is dependent on boost-velocity 
$\boldsymbol{v}$ as in Eq.~(\ref{2.28b}). 
Each parameter/factor appearing in Eq.(\ref{2.38}) 
takes the following forms, respectively, in PF and in 
OF, and is mutually related as 
\beq
\mbox{PF}& & \mbox{OF}\nonumber\\
\{ \tau(\boldsymbol{v}(P)), \boldsymbol{P}(\boldsymbol{v}), P_{0}(\boldsymbol{v})=E\left(\boldsymbol{P}(\boldsymbol{v})\right)\}&\stackrel{\boldsymbol{v}= \boldsymbol{0}}{\rightarrow}&
\{T, \boldsymbol{0}, M\}, \nonumber\\
e^{\pm i \boldsymbol{P}(\boldsymbol{v})\cdot\boldsymbol{X}}\cdot e^{\mp i P_{0}(\boldsymbol{v})\tau}
&\stackrel{\boldsymbol{v}= \boldsymbol{0}}{\rightarrow}&
1\cdot e^{\mp i M T}.
\eeq
Similarly the urciton spinor WF's{\footnote{
In Eq.~(\ref{215}) we have concerned only the relevant $SU(2)_{\mathfrak{m}}$ freedom, and 
retained only the suffix $r/\bar{r}$.
}} 
in PF, $W(\boldsymbol{v}(P))$, are transformed from those in OF 
$W(\boldsymbol{0})$ by Booster 
$S_{B}$(Eq.~(\ref{Sb})) as
\begin{subequations}
\label{215}
\beq
W_{r,\alpha}(\boldsymbol{v}(P))&\equiv&
\{ U_{r,\alpha}(\boldsymbol{v}), V_{\bar{r},\alpha}(\boldsymbol{v})\}
=[S_{B}\left(\boldsymbol{v}\left(P\right))W_{r}(\boldsymbol{0}\right)]_{\alpha}, \nonumber\\
W_{r,\alpha}(\boldsymbol{0}) &\equiv& \{ U_{r,\alpha}(\boldsymbol{0}), V_{\bar{r},\alpha}(\boldsymbol{0})\}, 
\label{215a}
\eeq
where the latter is determined from the static unitary-symmetry 
$SU(2)^{(q)}_{\mathfrak{m}}\otimes SU(2)^{(\bar{q})}_{\mathfrak{m}}$ as 
\beq
\label{215b}
U_{r,\alpha}(\boldsymbol{0}): \ \ U_{+}(\boldsymbol{0})_{,\alpha}
=\left(
\begin{array}{c}
\chi\\
0\\
\end{array} 
\right), \ \ 
U_{-}(\boldsymbol{0})_{,\alpha}
=\left(
\begin{array}{c}
0\\
\chi\\
\end{array} 
\right),
\ \ \ \ \nonumber\\
V_{\bar{r},\alpha}(\boldsymbol{0}):\ \ 
V_{+}(\boldsymbol{0})_{,\alpha}
=\left(
\begin{array}{c}
0\\
\chi\\
\end{array} 
\right), \ \ 
V_{-}(\boldsymbol{0})_{,\alpha}
=\left(
\begin{array}{c}
\chi\\
0\\
\end{array} 
\right).
\eeq
\end{subequations}
By inspection of the equations (\ref{2.38}) to (\ref{215}) it may be ascertained concretely how 
the static unitary symmetry $U(4)_{DS,\mathfrak{m}}$ 
at rest frame ($\boldsymbol{P}=0$, $\boldsymbol{v}=0$) 
is to be embedded  in the covariant $\widetilde{U}(4)_{DS,\mathfrak{m}}$ space. 
 Thus we see that $\Phi_{\boldsymbol{P},\alpha}(\boldsymbol{X},\tau(\boldsymbol{v}))$, 
Eq. (\ref{2.38}), 
in OF($\boldsymbol{v}=0$), which is a {\it certain inertial frame}, is equivalent 
to the WF of constituent quark in Eq.~(\ref{Eq10}) with $\boldsymbol{P}=0$, 
that is, 
\begin{subequations}
\label{2.16}
\beq
\label{2.16a}
\Phi_{\boldsymbol{P},\alpha}(\boldsymbol{X},\tau(\boldsymbol{v}))|_{\boldsymbol{v}=0}
=\Phi_{\alpha}(X_{\mu})|_{\boldsymbol{P=0}}, 
\eeq
and also that, considering in the whole inertial frame, the following 
two wave functions are mutually equivalent 
\beq
\label{2.16b}
&&\{ \Phi_{\bs{P}(\bs{v}),\alpha}(\boldsymbol{X},\tau(\boldsymbol{v}))|_{ \boldsymbol{v}={\rm whole}
(-1<v_{i}<+1)}\}\nonumber\\
&&=\{\Phi_{\alpha}(X_{\mu})|_{P_{\mu}^2=-m_{q}^2}; {\rm Constit. Q. WF~(Eq.(\ref{Eq10}))~in 
~Evolved~COQM}\}. 
\nonumber\\
\eeq
\end{subequations}
Here it is to be noted that in $\tau$-Quantum Mechanics there exists 
{\it no negative-energy problem}. Instead of the ``problem'', it implies 
an ``application of crossing rule'' in confined quark-level 
without violating the color-singlet condition of its parent-hadron. 
We also should like to remark that 
in the solution (Eq. (\ref{2.38})) of $\tau$--Schr{\"o}dinger 
wave equation (Eq. (\ref{2.36})) 
the chiral-urciton, whose {\it sole-existence} is prohibited by 
{\it Least-Action Principle}, appears in
Eq.~(\ref{2.38}) now in the confined system, being accompanied by 
the Pauli-urciton; This pair of urcitons plays a role of basic vectors of the chirality-symmetric 
$SU(2)_{\mathfrak{m}}$ space.  
\subsection{Prototype Mechanics for Multi-Urciton Quark Fields}
\label{2.2}
Then we shall study the prototype mechanics for Yukawa-type Klein-Gordon field, describing the 
composite hadrons of multi-urciton quarks in the 
$\widetilde{U}(4)_{DS, \mathfrak{m}}$-classification scheme. 

\subsubsection{Internal Coordinates and Proper time for 
Multi-Quark system}
\label{new3-2-1}

First of all we have to start with consideration on the notion of proper time for 
a multi-body classical-particle system. For the present purpose we are concerned only space-like 
internal extensions, and define the external/center of mass coordinate $X_{\mu}$ and the internal coordinate 
$r_{\mu}^{(i)}$ for each constituent, respectively, as 
\begin{subequations}
\label{54}
\beq
\label{54a}
X_{\mu}\equiv\frac{1}{M_0}\sum_{i} m_{i}x_{\mu}^{(i)}, \ x_{\mu}^{(i)}\equiv r_{\mu}^{(i)}+X_{\mu},
\eeq
\beq
\label{54b}
\sum_{i} m_{i}\bs{r}^{(i)} =0, 
\eeq
\end{subequations}
where $M_0 \equiv \sum_{i} m_{i} $ and $x_{\mu}^{(i)}$'s ($i=1$ to 
$n+m$; $n (m)$ is number of quarks (anti-quarks)) denote the 4-coordinate 
of respective constituents. 
It is to be noted that each internal coordinate $r_{\mu}^{(i)}$ is measured from 
its origin, the external coordinate $X_{\mu}$-itself, of which choice is critical for simple 
presentation of our theoretical framework. The point is as follows: 
Since the internal extension is space-like, it is possible to choose the Lorentz frame so that 
all the time variables of the external and of the respective urcitons are same, implying 
vanishing of all internal-time variables as, 
\beq
r_{0}^{(i)} \equiv x_{0}^{(i)}-X_{0}=T-T=0, {\mbox{or more generally}} \ P_{\mu}r_{\mu}^{(i)}=0.
\label{2.18}
\eeq
This means that, in the relevant Lorentz frame, 
the all internal coordinates become two-dimensional vectors as 
$r_{\mu}^{(i)}=(\boldsymbol{r}^{(i)},0)$, and satisfying $\boldsymbol{v}\cdot
\boldsymbol{r}^{(i)}=0$. 
In the case of $\widetilde{U}(4)_{DS, \mathfrak{m}}$-scheme, where all urcitons do perform the 
parton-like motion(see, Eq.(\ref{EqA9}) in Appendix \ref{AppendixA}), that 
all of their boost-velocity is equal to that of parent 
as $\boldsymbol{v}^{(i)} = \boldsymbol{v}$, 
the proper time of all constituents become identical to that of total system as 
\beq
d\tau (\boldsymbol{v}^{(i)})= d\tau (\boldsymbol{v}). 
\label{2015-2-3-1}
\eeq
Therefore, in our scheme, the troublesome {\it relative-time problem}, appearing in most 
of the relativistic composite models, disappears at the beginning. 
It is also to be noted that
\beq
\tau(\boldsymbol{v}) \ \mbox{in PF} 
\stackrel{\boldsymbol{v} \to 0}{=}T \ \mbox{in OF}. 
\label{particle-observer}
\eeq
Now we are possible to proceed to classical mechanics. 

\subsubsection{Classical Mechanics} 
\label{Classical Mechanics}

In classical mechanics the relevant action is given by 
\label{bilocal action}
\beq
S&=&\int_{a}^{b}{\mathcal{L}}
(\dot{X}_{\mu};\boldsymbol{r}^{(i)}\mbox{'s}, 
\boldsymbol{\dot{r}}^{(i)}\mbox{'s}) d\tau (\boldsymbol{v}), \nonumber \\ 
{\mathcal{L}}&=&-\sqrt{2U(\boldsymbol{r}^{(i)}\mbox{'s})}
\sqrt{-\left(M_0\dot{X}^2_{\mu}+\sum_{i} m_{i} \boldsymbol{\dot{r}}^{(i)2} \right)}, 
\ \  U(\boldsymbol{r}^{(i)}\mbox{'s})\equiv \sum_{i}U^{(i)}(\bs{r}^{(i)})>0, 
\label{2.38a}
\eeq
in terms of $\dot{X}_{\mu}$,  
$\dot{\boldsymbol{r}}^{(i)}\mbox{'s}$, 
$\boldsymbol{r}^{(i)}\mbox{'s}$; 
and{
$d\tau=dT\sqrt{1-\boldsymbol{v}^2}$. 

As was mentioned{\cite{Ishida-Sonoda}} our Lagrangian (\ref{2.38a}) is not of additive form 
as regards constituent-particles, and does not reduce to the sum of respective ones 
even in the free limit with constant $U$. It is 
also notable that this form of Lagrangian makes possible the separation of motion into the 
external and the internal ones. 

From Eq.~(\ref{2.38a}) we derive the expression of respective momenta 
\begin{subequations}
\beq
P_{\mu}=\frac{\partial {\mathcal L}}{\partial \dot{X}_{\mu}}=
\frac{\sqrt{2U}M_0 \dot{X}_{\mu}}{\sqrt{-(M_0 \dot{X}_{\mu}^2+\sum_{i}m_{i}\dot{\boldsymbol{r}}^{(i)2})}}
\stackrel{\cdot}{=}M_0 \dot{X}_{\mu},
\eeq
\beq
\boldsymbol{p}^{(i)}=\frac{\partial {\mathcal L}}{\partial \dot{\boldsymbol{r}}^{(i)}}
=\frac{\sqrt{2U} m_{i} \dot{\boldsymbol{r}}^{(i)}}{\sqrt{-(M_0 \dot{X}_{\mu}^2+\sum_{i} m_{i}\dot{\boldsymbol{r}}^{(i)2})}}
\stackrel{\cdot}{=} m_{i} \dot{\boldsymbol{r}}^{(i)},
\eeq
\end{subequations}
where in the last equality is chosen the ``covariant gauge'', now with a 
{\it $\tau$-gauge function $G$} to be 
$G \equiv \frac{\sqrt{2U}}{\sqrt{-(M_0 \dot{X}_{\mu}^2+\sum m_{i}\dot{\boldsymbol{r}}^{(i)2})}}
=1$.  These internal momenta have to satisfy the ``Kinematical 
constraints''\footnote{\label{foot1}
This kinematic constraint comes from the definition Eq.~(\ref{54}) of coordinate $r_{\mu}^{(i)}$
's in the classical mechanics level, as $\frac{d}{d\tau}\sum_{i}m_{i}\bs{r}^{(i)}
=\sum_{i}m_{i}\dot{\bs{r}}^{(i)}=\sum_{i}\bs{p}^{(i)}=0$. 
} 
\beq
\label{60}
\sum_{i}\bs{p}^{(i)}=0. 
\eeq

Lagrange equations of motion are also derived in the particle frame, as 
\begin{subequations}
\beq
\frac{d}{d\tau}M_0 \dot{X}_{\mu}&=&\frac{d}{d\tau}P_{\mu}=0,
\label{2013-2-19-1}
\eeq
\beq
\frac{d}{d\tau}m_{i} 
\dot{\boldsymbol{r}}^{(i)}+\frac{\d U^{(i)}(\boldsymbol{r}^{(i)})}
{\d \boldsymbol{r}^{(i)}}
&=&\frac{d}{d\tau}\boldsymbol{p}^{(i)}
+\frac{\d U^{(i)}(\boldsymbol{r}^{(i)})}{\d \boldsymbol{r}^{(i)}}=0, 
\eeq
\end{subequations}
while in the observer frame, as
\begin{subequations}
\beq
\frac{d}{d T}P_{\mu}=0 \ \ \ (\boldsymbol{P}=0, \ P_{0}=M_0), 
\eeq
\beq
\frac{d}{d T}\boldsymbol{p}^{(i)}+\frac{\d U^{(i)}(\boldsymbol{r}^{(i)})}{\d \boldsymbol{r}^{(i)}}=0. 
\eeq
\end{subequations}

\paragraph{({\it $\tau$-Hamiltonian and Hooke Potential})}

Then Hamiltonian ${\mathcal H}$ is given(, corresponding to Eq.~(\ref{2.34}) in the case of local urciton field,) by the definition, as 
\begin{subequations}
\label{2.40}
\beq
{\mathcal H}&\equiv&P_{\mu} \dot{X}_{\mu}+ \sum_{i}\boldsymbol{p}^{(i)}\cdot\dot{\boldsymbol{r}}^{(i)}-
{\mathcal{L}}(\dot{X}_{\mu}; \boldsymbol{\dot{r}}^{(i)}\mbox{'s}, \boldsymbol{{r}}^{(i)}\mbox{'s})
=\frac{1}{2M_0}(P_{\mu}^2+{\mathcal M}^2(\boldsymbol{{r}}^{(i)}\mbox{'s}, \boldsymbol{{p}}^{(i)}\mbox{'s}))
=0, \nonumber\\
\label{2.40a}
\eeq
\beq
\label{2.40b}
{\mathcal M}^2=\sum_{i} \mathfrak{m}^{(i)2}
(\boldsymbol{{p}}^{(i)},\boldsymbol{{r}}^{(i)}),  \ \ 
\mathfrak{m}^{(i)2}(\boldsymbol{{p}}^{(i)},\boldsymbol{{r}}^{(i)})
\equiv \frac{M_0}{m_{i}}{\boldsymbol{p}}^{(i)2}+2M_0 U^{(i)}(\boldsymbol{{r}}^{(i)}).
\eeq
\end{subequations}
Here {\it zero} in the last equality in Eq.~(\ref{2.40a}) 
comes from the $\tau$-gauge invariance of action (\ref{2.38a}). 
As a concrete form of potential $U^{(i)}(\bs{r}^{(i)})$ 
in Eq.~(\ref{2.40b}), 
we apply, as a simple model the one-dimensional Hooke's one (\ref{U-pot-a}), 
which, due to Eqs.~(\ref{2013-2-19-1}) through (\ref{U-pot-b}), 
leads the respective 
mass-operator $\mathfrak{m}^{(i)2}$ (\ref{2.40b})
to an oscillator (\ref{U-pot-b}) for deviation of internal extension 
$\boldsymbol{r}^{(i)}$'s from its average position $\boldsymbol{r}^{(i)}=0$
(that is, $\boldsymbol{X}$) as 
\begin{subequations}
\label{U-pot}
\beq
U^{(i)}(\boldsymbol{r}^{(i)})=
U^{(i)}(0)+\frac{1}{2}K^{(i)}\boldsymbol{r}^{(i)2}, 
\label{U-pot-a}
\eeq
\beq
\mathfrak{m}^{(i)2}(\boldsymbol{p}^{(i)},\boldsymbol{r}^{(i)})
 =2M_0 U^{(i)}(0)
+M_0 [(\bs{p}^{(i)2}/m_{i})+K^{(i)}\bs{r}^{(i)2}].
\label{U-pot-b}
\eeq
\end{subequations}
Thus far in this subsection \ref{Classical Mechanics}, 
we have shown our action Eq.~(\ref{2.38a}) describes surely 
such a property of confined multi-quark system that, total mass 
$M$ is conserved, and each constituent makes independently a 
simple harmonic oscillation around their center of mass position. 

\subsubsection{Quantum Mechanics -- the First-Stage Quantization}
\label{Quantum Mechanics}
\paragraph{({\it 1st Quantization})} 

Then our relevant 1st-quantization may be performed by replacing the 
P.B. with commutator 
between the canonical variables 
as 
\beq
\{X_{\mu}, P_{\nu}\}_{\rm P.B.}=\delta_{\mu\nu} &\to& 
\left[\hat{X}_{\mu}, \hat{P}_{\nu}\right]_{\rm com.}=i \delta_{\mu\nu}, \nonumber \\
\{r_{k}^{(i)}, p_{l}^{(i)}\}_{\rm P.B.}=\delta_{kl}^{(i)} &\to& 
\left[\hat{r}_{k}^{(i)}, \hat{p}_{l}^{(i)}\right]_{\rm com.}=i \delta_{kl}^{(i)},
\label{2.41a}
\eeq 
and by setting the ``Proper-time'' Schr{\"o}dinger 
Wave Equation (WE). 

\paragraph{({\it $\tau$-Schr{\"o}dinger Equation and Gauge Constraint})}

With $\tau$-Hamiltonian (\ref{2.40a}), $\tau$-Schr{\"o}dinger WE is 
written in PF as 
\beq
\label{2.41}
&&i\frac{d}{d\tau}\Phi_{\bs{P}, \alpha_{1}\cdots\alpha_{n}}^{\beta_{1}\cdots\beta_{m}}(\boldsymbol{X},\boldsymbol{r}^{(i)}\mbox{'s}:\tau)
=\hat{\mathcal H}(\hat{\boldsymbol{P}},\hat{\boldsymbol{p}}^{(i)}\mbox{'s},\boldsymbol{r}^{(i)}\mbox{'s}:\tau)
\Phi_{\bs{P}, \alpha_{1}\cdots\alpha_{n}}^{\beta_{1}\cdots\beta_{m}} (\boldsymbol{X},\boldsymbol{r}^{(i)}\mbox{'s}: \tau), 
\nonumber\\
&&\hat{\mathcal{H}}\equiv \frac{1}{2M_0}\hat{G}=
\frac{1}{2M_0}(\hat{P}_{\mu}^2+\hat{\mathcal{M}}^2).
\label{tau-H}
\eeq 
On the other hand the gauge constraint, the last equality 
of Eq.~(\ref{2.40a}), ( corresponding to 
Eq.({\ref{2.37}}) in the local case, ) is 
expressed{\footnote{
See the footnote on Eq.~(\ref{2.37}).
}} in OF as 
\beq
\label{2.42}
&&\hat{G}\Phi_{\alpha_{1}\cdots\alpha_{n}}^{\beta_{1}\cdots\beta_{m}}(X_{\mu},\boldsymbol{r}^{(i)}\mbox{'s})=
\left(\hat{P}_{\mu}^2+{\hat{\mathcal{M}}^2(\boldsymbol{\hat{p}}_{}^{(i)}\mbox{'s},
\boldsymbol{r}^{(i)}
\mbox{'s})}\right)
\Phi_{\alpha_{1}\cdots\alpha_{n}}^{\beta_{1}\cdots\beta_{m}}(X_{\mu},\boldsymbol{r}^{(i)}\mbox{'s})=0,  
\label{v8-68}
\eeq
\begin{subequations}
\beq
\hat{\mathcal{M}}^2=\sum_{i}
\hat{\mathfrak{m}}^{(i)2}(\hat{\bs{p}}^{(i)},\bs{r}^{(i)}),  \ \ \ 
\hat{\mathfrak{m}}^{(i)2}=
M_0 \left(\frac{1}{m_{i}}\hat{\boldsymbol{p}}^{(i)2}+2U^{(i)}(\boldsymbol{r}^{(i)})\right),
\eeq
\beq
\hat{\mathcal{M}}^2\equiv M_0 \hat{\mathcal{M}}, 
\label{4-12-4}
\eeq
\end{subequations}
which is the right Yukawa-type Klein-Gordon equation for Multi-Local 
Hadron-field in the {\it evolved-classification scheme}.
In the equations (\ref{2.41}) and (\ref{2.42}) we have simply put, similarly as 
Eqs.~(\ref{2.36}) and (\ref{2.37}) in the local case, 
the spinor indices on the $\tau$-WF. This may be allowed from the same 
reason as explained on the $\tau$-WF in the local 
case (see, Eqs.~(\ref{2.38}) through (\ref{215}) and 
the footnote \ref{foot14}). The only difference 
in the two cases is that the objects of application are 
basic vectors and their tensors of $SU(2)_{\mathfrak{m}}$-spin group, respectively, in the 
local and in the multi-local case. 
Here it is to be noted that the solution of Eq.(\ref{2.42}) plays the role of 
Wave Functions of the {\it Evolved COQM} with chirality-symmetric 
constituent quarks, which was proclaimed at the beginning of 
\S \ref{sec1.2.2}. 

By factorizing{\footnote{See the implication i) in Footnote \ref{footnote26}).}}
the gauge operator $\hat{G}$ in (\ref{2.42}) 
into the positive- and negative-frequency parts, $\{\Phi^{(+)},\Phi^{(-)}\}$, 
the $\tau$-Schr{\"o}dinger wave equation 
(\ref{2.41}) leads to the following equations (\ref{69}), as 
\beq
\label{68}
\hat{G}=(-\hat{P}_{0}+\hat{\xi})(\hat{P}_{0}+\hat{\xi}), \ 
\hat{\xi}\equiv\sqrt{\hat{\bs{P}}^2+\hat{\mathcal{M}^2}}; 
\eeq
\begin{subequations}
\label{69}
\beq
\label{69a}
i\frac{d}{d\tau}\Phi_{\bs{P},\alpha_{1}\cdots\alpha_{n}}^{(\pm)\beta_{1}\cdots\beta_{m}}
(\bs{X},\bs{r}^{(i)}\mbox{'s};\tau)=\pm \hat{\xi} 
\Phi_{\bs{P},\alpha_{1}\cdots\alpha_{n}}^{(\pm)\beta_{1}\cdots\beta_{m}}
(\bs{X},\bs{r}^{(i)}\mbox{'s};\tau) \ \mbox{in PF},
\eeq
\beq
\label{69b}
i\frac{d}{dT}\Phi_{{\rm (OF)},\alpha_{1}\cdots\alpha_{n}}^{(\pm)\beta_{1}\cdots\beta_{m}}
(\bs{r}^{(i)}\mbox{'s};T)=\pm \hat{\mathcal{M}}
\Phi_{{\rm (OF)},\alpha_{1}\cdots\alpha_{n}}^{(\pm)\beta_{1}\cdots\beta_{m}}
(\bs{r}^{(i)}\mbox{'s};T) \ \mbox{in OF}.
\eeq
\end{subequations}
These equations (\ref{69a}), (\ref{69b}) imply the crossing rule for composite hadrons 
with substitution of WF's, $\Phi^{(H/\bar{H})}\equiv \Phi^{(+/-)}$
(, corresponding to 
Eq.(\ref{2.38}) in the case of solitary Urciton-Quark.) 
Here, it may be worthwhile to remark that the $\tau$-Hamiltonian 
$\hat{\mathcal{H}}$ in (\ref{tau-H}), plays a role of ``{\it Prime-Hamiltonian}'', while 
the mass-spectral operator $\hat{\mathcal{M}}$ in Eq. (\ref{69b}) does the 
role of ``{\it real~}'' Hamiltonian in ordinary Quantum Mechanics 
 
\paragraph{({\it Concrete Form of $\tau$-Wave Function and mass operator in 
Non-Local One-Urciton Case})}

For instruction we insert here a detailed form of the solution 
with definite momenta $\bs{P}(\bs{v})$, for a couple of Eqs.~(\ref{2.41}) and 
(\ref{2.42}) in the ideal case of one-urciton system (, obtained by 
the aid of Eq.~(\ref{69})) as 
\beq
&&\Phi_{\bs{P},\alpha}(\bs{X},\bs{r};\tau(\bs{v}))
=\Phi^{(+)}_{\bs{P},\alpha}(\bs{X},\bs{r};\tau(\bs{v}))+\Phi^{(-)}_{\bs{P},\alpha}(\bs{X},\bs{r};\tau(\bs{v}))\nonumber
\\
&=&
\left\{
         \begin{array}{c}
\left(
\displaystyle{\sum_{r,s}}U_{r,s;\alpha}(P,M)e^{i\bs{P}(\bs{v})\cdot\bs{X}-i\tau(\bs{v})\hat{\xi}}
+
\displaystyle{\sum_{\bar{r},\bar{s}}}V_{\bar{r},\bar{s};\alpha}(-P,M)e^{-i\bs{P}(\bs{v})\cdot\bs{X}+i\tau(\bs{v})\hat{\xi}}\right) 
O_N (\bs{r})_{\bs{r}\perp\bs{v}} \\ \\
\left(
\displaystyle{\sum_{r,s}}U_{r,s;\alpha}(\bs{v}(P))e^{i\bs{P}(\bs{v})\cdot\bs{X}-i\tau(\bs{v})\hat{\xi}}
+
\displaystyle{\sum_{\bar{r},\bar{s}}}V_{\bar{r},\bar{s};\alpha}
(-\bs{v}(P))e^{-i\bs{P}(\bs{v})\cdot\bs{X}+i\tau(\bs{v})\hat{\xi}}
\right) 
O_N (\bs{r})_{\bs{r}\perp\bs{v}} ,
\end{array}
\right.
\label{70a}
\eeq
\beq
\label{70d}
\hat{\mathcal{M}}^2=\hat{\mathfrak{m}}^2(\bs{p},\bs{r})=
\hat{\bs{p}}^2+2mU(\bs{r}), 
\eeq
where $\hat{\xi}$ is defined in Eq.~(\ref{68}), $O_N (\bs{r}) $ is a set of 
oscillator functions relevant to mass operator (\ref{70d}), 
and $\{U(P,M)$/ $U(\bs{v}(P)), $
$V(-P,M)$/$V(-\bs{v}(P))\}$ in 
Eq.(\ref{70a}), respectively, 
are the basic Dirac-spinors for constituent/Urciton quarks in 
$\widetilde{U}(4)_{DS,\mathfrak{m}}$-spinor space defined in Eqs.(\ref{Eq11a}) and (\ref{Eq11b}), 
respectively. 
The above solution, corresponding to 
{\it local Urciton-Quark WF} Eq.~(\ref{2.38}),  clearly shows 
how to separate {\it the internal $\bs{r}$-space $O(2)_{\bs{r}\perp\bs{v}}$ 
from the $O(3,1)_{\rm Lor.}$-space} and how the static unitary symmetry $U(4)_{DS,
\mathfrak{m}}\otimes O(2)_{\bs{r}\perp\bs{v}}$ is embedded in the covariant 
refined-classification scheme $\widetilde{U}(4)_{DS,\mathfrak{m}}\otimes
O(3,1)_{\rm Lor}$. 
In the case of applying Hooke's potential (Eq.(\ref{U-pot})) the 
$\hat{\mathfrak{m}}^2$ becomes 
\beq
\label{Hooke}
\hat{\mathfrak{m}}^2=2mU(0)\delta^2 (\bs{r})+\hat{N}{\Omega}
.\eeq
Here 
\beq
\hat{N} {\Omega}=\hat{\bs{p}}^2+mK\bs{r}^2 \ (\hat{N}\equiv
a_{}^{\dagger}a_{}; {\rm Number \ Operator}, {\Omega}=2m\sqrt{K/m}).
\eeq
As is well known 
the 1st/2nd terms in both the 
two equations (\ref{U-pot-a}) and (\ref{U-pot-b}) contribute, respectively, 
to the ground $S$-wave/excited states, 
and the $\hat{\mathfrak{m}}^{2}$ is rewritten as 
\begin{subequations}
\label{2013-2-19-2}
\beq
\hat{\mathfrak{m}}^2=m^2+\hat{N} {\Omega}.
\label{2013-2-19-2-a}
\eeq
The 1st term of this equation is derived as a result of the 
relation, as 
\beq
\label{v24-2.27}
m_{}=2 {U}^{}(\boldsymbol{0}).
\eeq
\end{subequations}

Thus we see the physical situation{\footnote{
It is to be noted that the physical 
content of Eq.~(\ref{v24-2.27}) does not contain 
any effect from something like {\it zero-point oscillation}. This 
comes from our choice of origin of internal coordinate $\boldsymbol{r}^{(i)}$'s, 
being $\boldsymbol{X}$(external/center of mass coordinate), and the 
oscillators on $\boldsymbol{r}^{(i)}$'s 
are pure {\it C-number}.}}
that the constituent quark mass $m_{i}$ is produced by an average value of 
the potential $U(\bs{r})$ at its origin. 
Here it may be notable that the $\mathcal{L}$ (Eq.~(\ref{2.38a})) in one-urciton 
case, so far as concerned with the ground state, becomes owing to 
this relation  identical to the 
$\mathcal{L}$ (Eq.~(\ref{2.27})) for a solitary urciton-quark. 

\paragraph{({\it Concrete Form of Multi-Urciton $\tau$-Wave Function and Its Relation to Multi-Quark WF in 
$\widetilde{U}(4)_{DS,\mathfrak{m}}\otimes O(2)_{\bs{r}\perp\bs{v}}$-Classification Scheme})}

In the preceding sub-section \ref{QMsolo1}, we have described in detail 
the structure of Urciton-Quark WF, 
$\Phi_{\bs{P},\alpha}(\boldsymbol{X},\tau (\boldsymbol{v}))$ in 
a certain inertial frame with definite boost velocity $\boldsymbol{v}$; 
and shown that its whole entity for all possible inertial frame 
with($-1<v_{i}<+1$) is equivalent to $\Phi_{\alpha} (X_{\mu})$, 
that is the constituent-quark WF in the 
$\widetilde{U}(4)_{DS,\mathfrak{m}}$-spin scheme. 

Further, through all the contents 
in the first half of this subsection \ref{2.2} and the discussions around 
Eq.~(\ref{2.18}), we have seen that the 
freedoms on {\it internal} coordinates $\boldsymbol{r}^{(i)}$'s are disconnected with 
that on 
External/center of mass coordinates $\bs{X}$. 

Therefore our relevant relation is derived, 
extending directly the equation(\ref{2.16b}) in the case of solitary Urciton quark, as 
\beq
\label{Multi-2.16b}
&&\{ \Phi^{\beta_{1}\cdots\beta_{m}}_{\bs{P}(\bs{v}), \alpha_{1}\cdots\alpha_{n}}(\boldsymbol{X}, 
\bs{r}^{(i)}{\rm 's}; \tau(\boldsymbol{v}))|_{ \boldsymbol{v}={\rm whole}
(-1<v_{i}<+1)}\} \nonumber \\
&&=\{\Phi^{\beta_{1}\cdots\beta_{m}}_{\alpha_{1}\cdots\alpha_{n}}(X_{\mu}, \bs{r}^{(i)}{\rm 's});{\rm Multi \ Q. \ WF \ in \ Evolved \ COQM} \ \rm Eq.~(\ref{2.42})\}. 
\nonumber\\
\eeq

\paragraph{({\it Non-local Multi-Quark Wave Function - 
It's Expansion into Local $\widetilde{U}(4)_{DS,\mathfrak{m}}$-Spin Multiplets})}
 
For later use Basic Formulas in applying $\tau$-Quantum Mechanics to physical 
phenomena, obtained based upon the relation (\ref{Multi-2.16b}), 
are collected here. 
Firstly Multi-Quark WF 's are expanded, 
in terms of the oscillator functions $O_{\hat{N}}(\boldsymbol{r}^{(i)}\mbox{'s})$, 
into the Local WF's of $\widetilde{U}(4)_{DS,\mathfrak{m}}$
($\equiv \widetilde{U}(4)_{DS,\mathfrak{m}}^{(q)}\otimes
\widetilde{U}(4)_{DS,\mathfrak{m}}^{(\bar{q})}$)-multiplets 
$\Phi_{\hat{N}}^{(H)}(X)$ and $\Phi_{\hat{N}}^{(\bar{H})}(X)$ with respective 
quantums $\hat{N}$; as 
\begin{subequations}
\label{5-29-9}
\beq
\Phi_{\alpha_{1}\cdots\alpha_{n}}^{(H)\beta_{1}\cdots\beta_{m}}(X, 
\bs{r}^{(i)}{\rm 's})
&=&\sum_{\hat{N}} \Phi_{\hat{N},\alpha_{1}\cdots\alpha_{n}}^{(H)\beta_{1}\cdots\beta_{m}}(X)
O_{\hat{N}}(\boldsymbol{r}^{(i)}\mbox{'s}), \ \nonumber\\
\Phi_{\beta_{1}\cdots\beta_{m}}^{(\bar{H})\alpha_{1}\cdots\alpha_{n}}(X, 
\bs{r}^{(i)}{\rm 's})
&=&\bar{\Phi}_{\alpha_{1}\cdots\alpha_{n}}^{(H)\beta_{1}\cdots\beta_{m}}(X, 
\bs{r}^{(i)}{\rm 's}). 
\label{5-29-1}
\eeq
From these formulas, we mean $X \rightarrow X_\mu$. 
Then they are Fourier expanded as 
\beq
\label{5-29-2}
&&\Phi_{\hat{N},\alpha_{1}\cdots\alpha_{n}}^{(H)\beta_{1}\cdots\beta_{m}}(X)
=\sum_{P_{N}(P_{N,0}>0)}(
W_{\hat{N},\alpha_{1}\cdots\alpha_{n}}^{(H)\beta_{1}\cdots\beta_{m}}(\boldsymbol{v}(P_{N}))
e^{iP_{N}\cdot X}
+
W_{\hat{N},\alpha_{1}\cdots\alpha_{n}}^{(\bar{H})\beta_{1}\cdots\beta_{m}}(\boldsymbol{v}(P_{N}))
e^{-iP_{N}\cdot X}
), \nonumber\\
&&\Phi_{\beta_{1}\cdots\beta_{m}}^{(\bar{H})\alpha_{1}\cdots\alpha_{n}}(X)
=\bar{\Phi}_{\alpha_{1}\cdots\alpha_{n}}^{(H)\beta_{1}\cdots\beta_{m}}(X). 
\eeq
\end{subequations}
In Eqs.~(\ref{5-29-1}) and (\ref{5-29-2}) 
the internal space WF's and the spin WF's are, respectively defined, as:  
\begin{subequations}
\label{5-29-10}
\\
~~Concerning the Internal-Space WF{\footnote{
Here it is to be noted that the quantum number $l^{(i)}$, 
concerning 
internal orbital motion, {\it is missing} in the formulas Eqs.(\ref{2015-4-7-1}); 
This is a remarkable feature of our scheme (based on the $\tau$-Quantum Mechanics), 
in contrast to the conventional scheme (based on bound state picture) with 
$LS$-coupling force potential. See also, the item v) in section \ref{sec3.2}, and 
footnote \ref{foot29}.  See also the argument in \S \ref{sec0}.
}}
\beq
&&\hat{\mathcal M}^2 (\boldsymbol{p}^{(i)}\mbox{'s},\boldsymbol{r}^{(i)}\mbox{'s})
O_{\hat{N}}(\boldsymbol{r}^{(i)}\mbox{'s})
=M_{N}^2
O_{\hat{N}}(\boldsymbol{r}^{(i)}\mbox{'s}), \ \ 
M_{N}^2=\sum_{i} {m}^{2}_{N_{i}}, \nonumber\\
&&{\rm where}~\hat{\mathcal M}^2 (\boldsymbol{p}^{(i)}\mbox{'s},\boldsymbol{r}^{(i)}\mbox{'s})
=\sum_{i} \hat{m}^{(i)2}({\boldsymbol{p}^{(i)},\boldsymbol{r}^{(i)}})=
\sum_{i}\left(m_i^2+\hat{N}_{i}\Omega^{(i)}\right), {\rm and }~~\nonumber\\
&&O_{\hat{N}}(\boldsymbol{r}^{(i)}\mbox{'s})
=\prod_{i}O_{\hat{N}_{i}}(\boldsymbol{r}^{(i)});
~\hat{\mathfrak m}^2(\boldsymbol{r}^{(i)})O_{\hat{N}_{i}}(\boldsymbol{r}^{(i)})
={{m}}_{N_{i}}^2 O_{\hat{N}_{i}}(\boldsymbol{r}^{(i)}), \ 
m_{N_{i}}^2=m_{i}^2+N_{i}\Omega^{(i)}, \nonumber\\
&&\Omega^{(i)}=2 M_0\sqrt{\frac{K^{(i)}}{m_i}}, ~ 
\hat{N}\equiv \{\hat{N_{1}},\hat{N}_{2},\cdots\}, 
\ \hat{N}_{i}\equiv \{ N_{i}\}, N_{i}=2n_{i} (n_i =0,1,2,\cdots).~ 
\label{2015-4-7-1}\eeq

Concerning Spin WF
\beq
W_{\alpha_{1}\cdots}^{(H)\beta_{1}\cdots}
=\prod_{i, j}(
W_{\alpha_{i}}^{(q)}(\boldsymbol{v}(P))
\bar{W}^{(\bar{q}),\beta_{i}}(\boldsymbol{v}(P))
), \ 
W^{(\bar{H})\alpha_{1}\cdots}_{\beta_{1}\cdots}=
\overline{W^{(H)}}^{\beta_{1}\cdots}_{\alpha_{1}\cdots};\nonumber\\
\left[ (i u_{\mu}\gamma_{\mu}^{(i)}+\rho_{(m),3})W_{r}^{(q)}(\boldsymbol{v})
\right]_{\alpha_{i}}=0, \ \ 
\left[ W_{\bar{r}}^{(\bar{q})}(\boldsymbol{v})(iu_{\mu}\gamma_{\mu}^{(j)}+\bar{\rho}_{(m),3})
\right]^{\beta_{j}}
=0.
\label{2015-06-16-1}
\eeq
\end{subequations}
\subsubsection{Quantization of Local Multi-Urciton Hadron Fields -- the Second-Stage Quantization}\label{sec2.3.2}

\paragraph{({\it Covariant Quark-Representation of General Elementary Hadrons in the 
$\widetilde{U}(4)_{DS,\mathfrak{m}}$-Classification Scheme.}) }

The local WF's of $\widetilde{U}(4)_{DS,\mathfrak{m}}$-multiplet, 
$\Phi_{\hat{N}}^{(H)}(X)$/$\Phi_{\hat{N}}^{(\bar{H})}(X)$ defined in 
Eqs.(\ref{5-29-1}) and (\ref{5-29-2}) are, further decomposed into the members of multiplet, 
representing local hadrons $\phi^{(i)}(X)$/$\bar{\phi}^{(i)}(X)$ with definite chirality 
structure $\{\chi_{\hat{N}}(n, m)\}$/$\{\bar{\chi}_{\hat{N}}(n, m)\}$, as follows; 
\begin{subequations}
\label{2014-12-25-3}
\beq
\label{2014-12-25-3a}
\Phi_{\hat{N}(n, m);\alpha_{1}\cdots\alpha_{n}}^{(H)\beta_{1}\cdots\beta_{m}}(X)
&=&\sum_{i}\phi_{\chi_{\hat{N}}(n, m)}^{(i)}(X)
\Gamma_{\chi_{\hat{N}}(n, m);\alpha_{1}\cdots\alpha_{n}}^{(i)\beta_{1}\cdots\beta_{m}}, \\
\label{2014-12-25-3b}
\Phi_{\hat{N}(n, m);\beta_{1}\cdots\beta_{m}}^{(\bar{H})\alpha_{1}\cdots\alpha_{n}}(X)
&=&\sum_{i}\bar{\phi}_{\chi_{\hat{N}}(n, m)}^{(i)}(X)
\bar{\Gamma}_{\chi_{\hat{N}(n, m)}; 
\beta_{1}\cdots\beta_{m}}^{(i)\alpha_{1}\cdots\alpha_{n}},
\eeq
\beq
\vspace{-1em}
\label{2016-6-7-a}
\chi_{\hat{N}}\equiv ({}_{r_1,r_2, \cdots , r_{n}}^{\bar{r}_{1},\bar{r}_2, \cdots, \bar{r}_m}).
\eeq
\end{subequations}
This is an ideal form of covariant quark-representation of {\it elementary hadrons} 
$\phi^{(i)}(X)$/$\bar{\phi}^{(i)}(X)$ with definite chirality-structure $\{
\chi_{\hat{N}(n, m)}/\bar{\chi}_{\hat{N}(n, m)}\}$. 
In Eq.(\ref{2014-12-25-3}) the $\{\Gamma^{(i)}{\rm 's}/\bar{\Gamma}^{(i)}{\rm 's}\}$ are 
the tensor product of urciton-spinors concerned with respective constituent 
quarks, and plays a role of {\it Clebsch-Gordon Coefficients } to decompose the reducible WF, 
$\Phi^{(H)}_{\hat{N}}(X)/\Phi^{(\bar{H})}_{\hat{N}}(X)$, into the irreducible WF's for 
their members of elementary hadrons 
$\phi^{(i)}(X)/\bar{\phi}^{(i)}(X)$ with definite attributes $J^{PC\{\chi_{\hat{N}(n,m)}\}}$. 
Herewith it is to be noted that the $\{\Gamma^{(i)}{\rm 's}/\bar{\Gamma}^{(i)}{\rm 's}\}$, 
corresponding to WF in most of  
composite models based on bound-state picture, still has definite Lorentz-transformation 
property. 

Here it may be instructive to note that the other attributes $J^{PC}$, than the chirality 
$\chi_{\hat{N}}$ Eq.(\ref{2016-6-7-a}) of elementary hadrons, 
$\phi^{(i)}(X)/\bar{\phi}^{(i)}(X)$, are determined directly 
by the symmetry structure, Eq.(\ref{Eq6}), of WF, 
$\Phi_{\hat{N}}^{(H)}(X)$/$\Phi_{\hat{N}}^{(\bar{H})}(X)$, 
and the tensor property of the WF's, $\{\Gamma^{(i)}{\rm 's}/\bar{\Gamma}^{(i)}{\rm 's}\}$ 
for respective elementary hadrons. 

A concrete, simple example of the decomposition of 
$\widetilde{U}(4)_{DS,\mathfrak{m}}$-multiplet into 
its members with all the attributes has been given in the 
case of local $q\bar{q}$-meson field by Eqs.{(\ref{Eq17})} through (\ref{Eq18}) in 
\S \ref{sec1.2.2}. 
Then, they are to be quantized, 
following to the conventional procedure 
with the sacred ``Spin-Statistics connection'', strictly, as the 
2nd-stage quantization in our proper-time quantum mechanics.  

\paragraph{({\it Phenomenological Features of Evolved Classification Scheme on Light-Flavored 
Mesons and Baryons})} 

Application of the Evolved Hadron-Classification Scheme actually, 
taking into account of Flavor structure, 
to the relevant problem had been made, in detail, by the authors 
of Refs.\cite{Ishida2000} and 
\cite{Ishida2002}, of which framework is the base of Evolved Classification Scheme. 
(Here it should be noted that the terminology, ``$\widetilde{U}(12) \supset 
\widetilde{U}(4)_{DS}\otimes SU(3)_{F}$'' in Ref.\cite{Ishida2002} is identical to that 
``$\widetilde{U}(12)_{\mathfrak{m}} \approx 
\widetilde{U}(4)_{DS,\mathfrak{m}}\otimes SU(3)_{F}$'' in this work.)
Here, out of the results presented there, we shall pick up a few notable 
points as follows: 

i) In Evolved Classification scheme the spinor-WF’s of composite hadrons 
for $q\bar{q}$-mesons and $qqq$-baryons are given as tensor products of 
the basic spinors in the $\widetilde{U}(4)_{DS,\mathfrak{m}}$-spin 
space, which are the spinor WF of constituent quarks. The basic-spinors 
consist of Paulon / Chiralon with $J^{P}={\frac{1}{2}}^{+} / {\frac{1}{2}}^{-}$ 
(, see Eqs. (\ref{Eq3}) and (\ref{Eq4})), where the Chiralon produces 
systematically 
the seats of elementary composite-mesons / baryons with {\it new type of 
exotic-quantum numbers}, which would never appear in Non-relativistic scheme.  

ii) The spinor-WF of $q\bar{q}$ meson in the ground $S$-wave state is 
given in the form as $S^{(N/E)B}_{A} \approx 
(U_{r=+}\bar{V}_{\bar{r}=-}\pm U_{r=-}\bar{V}_{\bar{r}=+})_{a}^{b}
\cdot \delta_{i}{}^{j}$
 and $V^{(N/E)B}_{A} \approx 
(U_{r=+}\bar{V}_{\bar{r}=+}\pm U_{r=-}\bar{V}_{\bar{r}=-})_{a}^{b}
\cdot \boldsymbol{\sigma}_{i}{}^{j}$, 
respectively, for the $U(3)_F$-nonet Scalar Meson Field, $S^{(N/E)B}_{A}(
\boldsymbol{r})$, 
and for the $U(3)_F$-nonet Vector Meson Field $V^{(N/E)B}_{A}(\boldsymbol{r})$ 
(; See Table 1 in Ref.{\cite{Ishida2000}} and Table 2 in Ref.{\cite{Ishida2002}} ). 

Here it may be instructive to note that these expressions on $S^{(N/E)B}_{A}(\boldsymbol{r})$ 
and $V^{(N/E)B}_{A}(\boldsymbol{r})$ concern their dependence on the 
internal space-coordinate $(\boldsymbol{r})$ at a certain inertial frame ($\boldsymbol{v}=0$), 
while they {\it do} become, in the scope (Eq.(\ref{Multi-2.16b})) of all possible 
$\{\boldsymbol{v}{\rm 's}\}$, respectively, their fields as $\{S(X)\}$ and $\{V_{\mu}(X)~~
{\rm with~~
\partial_\mu V_\mu (X)=0}\}$.

Herewith, phenomenologically, the following two points are notable: 
Firstly the existence of seats for {\it light-mass} (, since of the ground state, ) scalar 
mesons, possibly corresponding to the $\sigma$-meson nonet-members \{ $\sigma (500)$, 
$\kappa(800)$, $a_0 (980)$, $f_0 (980)$\}. 
Secondly the existence of seats for extra low-mass vector-mesons 
\{ $\omega (\sim 1200)$, 
$\rho (1250)$, $\cdots$\}, now under the serious discussions theoretically and/or 
phenomenologically from various view points(, for example, see 
\S \ref{subsec4.2} and 
Ref.\cite{Komada}).

iii) The spinor-WF of ground $S$-wave state baryons, relevantly light-flavored, is 
classified into the three-type of chirality structure as, 
$E: U_{+}U_{+}U_{+}$, $F: U_{+}U_{-}U_{-}$ and $G: U_{+}U_{+}U_{-}$, of which 
$SU(6)_{SF}$ dimensions are, respectively, $\underline{56}$, $\underline{56}'$, 
and $\underline{70}$.  
Out of them the spinor-WF's of $\underline{56}'$ and $\underline{70}$ multiplets 
include the chiralon components and accordingly they contain the {\it exotic} 
elementary hadrons(, see, table 3 in Ref.\cite{Ishida2002}). 
Herewith phenomenologically, the following two points are notable: 
Firstly the {\it $\underline{56}'$-multiplet} includes the Roper-resonance 
$N(1440, 1/2^{+})$, which seemed too light to be a radially-excited $S$-wave 
state in the Non-relativistic scheme; while now in Evolved Scheme the mass 
difference between the two-ground $S$-wave states, $N(940, 1/2^{+})$ and 
$N(1440, 1/2^{+})$, 
may be understood as due to Broken-chirality by the effective strong QCD-interaction. 
(See, Item ``Chirality Structure of Condensing Vacuum-Pair WF'' and 
Footnote \ref{foot32} in section \ref{Degenerate Ground State}.)  
Secondly the $\underline{70}$-multiplet includes the $SU(3)_F$-singlet resonance 
$\Lambda (1405, 1/2^{-})$, which seemed 
too-light to be a $P$-wave state in the Non-Relativistic Scheme; 
while now to be 
of {\it reasonable mass}-value in Evolved Scheme by the same 
reason as in the case of Roper resonance. 

Herewith, it is to be noted that all the baryons with definite flavor-structure, 
as is listed in the PDG-booklet, is deserved to be called as {\it Elementally hadrons}, 
when their spinor-WF's are described in the chirality-representation. 
\subsubsection{Features and New aspects of Hadron Classification Scheme deduced from 
the $\tau$-Quantum Mechanics}
\label{sec2.2.4}

\paragraph{({\it Features of $\tau$-Quantum Mechanics}) }

Before entering into the specific topics, we firstly pick up general features of 
our first-stage $\tau$-Quantum Mechanics given in the preceding subsections \ref{2.1} and \ref{2.2}, as follows: 

\noindent a)~No ``Relative-Time'' Problem: In the most relativistic composite-models there exists the 
relative-time freedom, which generally leads to existence of serious unphysical-states. 
Contrarily,  in our case, we start from the classical system with a sole proper time through 
all constituents (see Eq.~(\ref{2015-2-3-1})). 

\noindent b)~Negative energy problem and Crossing Rule: In the most relativistic quantum field theories 
there appear{\footnote{
Note that the relevant WF has to satisfy Klein-Gordon wave equation with the second-order 
time derivative. 
}} 
negative-energy states, whose interpretation had been an annoying problem. 
For its disposal now generally it is applied so-called ``Crossing Rule'' rather 
{\it in ad hoc way.} However, this rule {\it be dictated},
{\footnote{
Here it may be instructive to note that development of the $\tau$ is always 
directed to Future (i.e. $d\tau >0$) and $\hat{\xi}$($\mathcal{M}$) has positive eigen-value. 
}}
in this work, as a result of the 
$\tau$-Schr{\"o}dinger wave equation, which contains only {\it first-order $\tau$-time derivative} 
(see, Eqs.~(\ref{69a}), (\ref{69b}) and also argument after Eq.~(\ref{2.16b}). ). 

\noindent c)~Origin of ``Unitary'' Quantum Numbers, as Attribute of Composite Hadron: 
Our wave functions of composite hadron are defined in 
$\widetilde{U}(4)^{(q)}_{DS,\mathfrak{m}}\otimes\widetilde{U}(4)^{(\bar{q})}_{DS,\mathfrak{m}}$-spinor 
$\otimes O(2)_{\bs{r}\perp\bs{v}}$-space; where its former part is, 
in OF, Lorentz-covariant; and embedded, in PF, the unitary symmetry 
${U}(4)^{(q)}_{DS,\mathfrak{m}}\otimes {U}(4)^{(\bar{q})}_{DS,\mathfrak{m}}$, 
which is the origin of separate-conservation of quark and anti-quark numbers etc.; 
while its latter part $O(2)_{\bs{r}\perp\bs{v}}$
has an unitary invariant norm. 
Needles to mention that the tensor property in the 
$\widetilde{U}(4)_{DS,\mathfrak{m}}$-space 
of the former part leads to existence of plural different 
intrinsic-spin members within a relevant 
single $\widetilde{U}(4)_{DS,\mathfrak{m}}$-multiplet. 

\paragraph{({\it New Aspects of Yukawa-type Klein-Gordon Equation for 
Multi-Local Hadron Field
})
} 
  
 The $\tau$-Quantum Mechanics has started from a classical action(Eq.~(\ref{2.38a})) 
 for multi-quark system, which seems 
 to represent well the physical situations of quark confinement. The relevant Klein-Gordon Equation for 
 the WF in the $\widetilde{U}(4)_{DS,\mathfrak{m}}\otimes O(2)_{\bs{r}\perp\bs{v}}$ scheme corresponds 
in the $\tau$-Quantum Mechanics to the gauge constraint, reflecting that the time scale of confined system is unobservable
(, see argument around Eq.~(\ref{2.32})). Moreover, the $\tau$-Quantum Mechanics has also given the other 
new aspects 
on the Klein-Gordon Equation as follows: 
\begin{itemize}
\item On Basic Vector of $SU(2)_{\mathfrak{m}}$-spin space: It leads to 
the Chiralon($J^{P}=(1/2)^{-}$) as a ``shadow particle'' of the Paulon($J^{P}=(1/2)^{+}$) ). 
It may be regarded as a ``{\it symbolic entity}'' of Quark 
confinement-mechanism(see, Argument after Eq.~(\ref{2.16b})). 
\item On $\hat{\mathcal{M}^2}$ and Potential $U^{(i)}(\bs{r}^{(i)}{\rm 's})$ (See, Eq.~(\ref{v8-68})): 
$\hat{\mathcal{M}^2}$ is an independent sum of respective harmonic oscillators of 
each constituent, and constituent quark mass is given by average value $\bar{U}^{(i)}(0)$ around 
center of mass position of parent(, see Eq.~(\ref{v24-2.27})). 
\item On True Nature of Constituent Quark / Urciton: It is a 
{\it simulator} of Bounded Quark by Non-Perturbative QCD potential, and plays a role of 
{\it carrier} of the spinor-indices ($\alpha_{i}{\rm 's}$) and the internal extension 
variables ($\bs{r}^{(i)}{\rm 's}$). Accordingly the various viewpoints on statistics of quarks in 
1960's (for example, para-statistics of order 3 {\cite{Greenberg}} 
and white-Bose Quarks{\cite{IshidaYamada1987}}{\footnote{
See, especially, S. Ishida and M. Oda cited in this Ref.{\cite{IshidaYamada1987}}.
}}) become presently 
meaningless. In this connection it is to be noted that the origin of all $\bs{r}^{(i)}{\rm 's}$ is 
fixed at the external coordinate $\bs{X}$ of the relevant composite hadron, and hence necessarily let 
our framework be satisfied Cluster Property.{\footnote{
See, a textbook on ``Specific Quantum Mechanics'' by S. Kamefuchi and M. Omote\cite{Specific Quantum Mechanics}.
}}
\end{itemize}

\section{Chirality Symmetry in Composite-Hadron Physics}
\subsection{Degenerate Ground State in Urciton-Quark Pair System and Vacuum Condensation}
\label{Degenerate Ground State}

\paragraph{({\it Chirality Conservation in Hadron Physics and Basic Urciton-Quark 
Wave Function})} 

In QCD (more generally in the standard gauge theory) the interaction of quarks with 
gluons (relevant gauge bosons) is introduced through the minimal substitution of covariant 
derivative, applied on the kinetic term of relevant quarks in the Lorentz-invariant free 
quark-action. Therefore, as has been discussed in the subsection \ref{sec1.2.1}, 
the chirality (a new discrete symmetry) is, in the case without condensed-vacuum effects,  
conserved in QCD (and all the gauge theories) regardless of light- or heavy-mass of 
the relevant quarks. 
Accordingly in the strong interaction (, whose underling basic dynamics 
is QCD), and also in all the other effective interactions the chirality should be conserved. 
In order to satisfy this requirement in a simple way the Urciton-Quark WF has 
been expanded 
in terms of the Extended Dirac Spinors in the 
$\widetilde{U}(4)_{DS,\mathfrak{m}}$-Scheme as Eq.~(\ref{Eq10}). 

\paragraph{({\it Bi-local $\tau$-WF for Urciton-Quark Pair $(U_{r},\bar{V}_{\bar{r}})$ System and 
Degenerate Ground State}) }

Making use of the general formulas (Eqs.~(\ref{5-29-9}) and (\ref{5-29-10})) 
for a system of multi-urciton system given 
in sub-section \ref{Quantum Mechanics}, we can write down the necessary formulas in the 
present case of Urciton-Quark Pair 
system as: \\
\begin{subequations}

Non-local WF : 
\beq
\label{5-29-3}
&&\Phi^{(P)}_{\alpha}{}^{\beta}(X;\bs{r}_{1},\bs{r}_{2})=\sum_{\hat{N}} \Phi_{\hat{N},\alpha}^{(P)}{}^{\beta}(X)
O_{\hat{N}}(\bs{r}_{1},\bs{r}_{2}), \nonumber \\
&&\Phi_{\beta}^{(\bar{P})}{}^{\alpha}(X;\bs{r}_{1},\bs{r}_{2})=
\sum_{\hat{N}} \Phi_{\hat{N},\beta}^{(\bar{P})}{}^{\alpha}(X)
O_{\hat{N}}(\bs{r}_{1},\bs{r}_{2}),
\eeq

Local-Pair WF: 
\beq
\label{5-29-4}
\Phi^{(P)}_{\hat{N},\alpha}{}^{\beta}(X)&=&\sum_{P_{N}(P_{N,0}>0)}
(W_{\alpha}^{(P)\beta}(\bs{v}(P_{N}))e^{iP_{N}X}+
 W_{\alpha}^{(\bar{P})\beta}(\bs{v}(P_{N}))e^{-iP_{N}X}),\nonumber\\
\Phi^{(\bar{P})}_{\hat{N},\beta}{}^{\alpha}(X)&=&\overline{\Phi^{(P)}}_{\hat{N},\alpha}{}^{\beta}(X)\nonumber\\
&=&\sum_{P_{N}(P_{N,0}>0)}
(W_{\beta}^{(\bar{P})\alpha}(\bs{v}(P_{N}))e^{iP_{N}X}+
 W_{\beta}^{({P})\alpha}(\bs{v}(P_{N}))e^{-iP_{N}X}).\nonumber\\
\eeq
\end{subequations}
\begin{subequations}

Internal-Space WF{
\footnote{
As has been described in detail in the sub-section \ref{Quantum Mechanics}, 
the relevant internal WF is determined by ${\hat{\mathcal{M}}^2}$, which 
is an independent sum of the 
respective Harmonic 
Oscillators on each $\bs{r}^{(i)}$, being connected with the center of mass of parent's hadrons. 
As for details, see around Eqs.(\ref{2.42}) through (\ref{69}). 
}}   : 
\beq
\label{5-29-5}
&&\hat{\mathcal{M}}^2(\bs{p}^{(1)}, \bs{r}^{(1)}; \bs{p}^{(2)}, \bs{r}^{(2)})
O_{\hat{N}}(\bs{r}^{(1)},\bs{r}^{(2)})=
M^{2}_{\hat{N}}O_{\hat{N}}(\bs{r}_{1},\bs{r}_{2})~~ {\rm and} ~~\nonumber\\
&&M_{\hat{N}}^2
=M^{2}_{\hat{N}_{1}}+M^{2}_{\hat{N}_{2}}=m_{1}^2+m^2_{2}+
N_{1}\Omega^{(1)}+N_{2}\Omega^{(2)}, \nonumber\\
&&{\rm where} ~~\hat{\mathcal{M}}^2(\bs{p}^{(1)}, \bs{p}^{(2)}; \bs{r}^{(1)},\bs{r}^{(2)})
=\hat{\mathfrak{m}}_{1}^{2}(\bs{p}^{(1)}, \bs{r}^{(1)})
+\hat{\mathfrak{m}}_{2}^{2}(\bs{p}^{(2)}, \bs{r}^{(2)}), \nonumber\\
&&\hat{N}=\{\hat{N}_{1}, \hat{N}_{2}\}, \ \ \hat{N}_{1, 2}\equiv \{ n_{1,2} \}, 
{N}_{1, 2}=2n_{1,2} \ (n_{1,2}=0,1,2,\cdots), \nonumber\\
&&O_{\hat{N}}(\bs{r}^{(1)},\bs{r}^{(2)})=O_{\hat{N}_{1}}(\bs{r}^{(1)})
O_{\hat{N}_{2}}(\bs{r}^{(2)}), \ 
\hat{\mathfrak{m}}^{(1,2)2}=m_{1,2}^2+\hat{N}_{1,2}\Omega^{(1,2)}  \ . 
\eeq

Intrinsic-Spin WF{\footnote{
Here we use the constituent quark spinors, Eqs. (\ref{Eq11a}).
}}
: 
\beq
\label{5-29-6}
W_{\alpha}^{(P)}{}^{\beta}(P,M)&=&
\sum_{r,\bar{r}}U_{r,\alpha}(P,M)\bar{V}_{\bar{r}}^{\beta}(P,M), \ \ \ 
W_{\alpha}^{(\bar{P})}{}^{\beta}(P,M)=
\sum_{r,\bar{r}}V_{\bar{r},\alpha}(P,M)\bar{U}_{{r}}^{\beta}(P,M); \nonumber\\
W_{\beta}^{(\bar{P})}{}^{\alpha}(P,M)&=&\bar{W}_{\alpha}^{({P})}{}^{\beta}(P,M), \ \  \ 
W_{\beta}^{({P})}{}^{\alpha}(P,M)=W_{\alpha}^{({\bar{P}})}{}^{\beta}(P,M); \ \ \
\eeq

while the urciton-spinors satisfy
the equations, as 
\beq
[\left(i u_{\mu}\gamma_{\mu}+ \rho_{(m),3}\right) U_{r}(\bs{v}(P))]_{\alpha}=0,  \ 
[\bar{V}_{\bar{r}}(\bs{v}(P))\left(i u_{\mu}\gamma_{\mu}+\bar{\rho}_{(m),3} \right)]^{\beta} =0.
\label{5-29-7}
\eeq
\end{subequations}

Here it may be instructive to note that the $\Phi^{(P)\beta}_{\alpha}(X,\bs{r}_{1},\bs{r}_{2})$ 
(\ref{5-29-3}) is the solution of $\tau$-Shr{\"{o}}dinger Eq. (\ref{tau-H}) with ``Prime''-Hamiltonian, 
while its positive- and negative- frequency parts (determined by the aid of Eq. (\ref{5-29-4})) 
correspond to 
those of ``{\it real} ''  $\tau$-Shr{\"{o}}dinger Eq. (\ref{69b}) with $\hat{\mathcal{M}}$, to be called as 
``{\it mass-spectral} '' Hamiltonian (See the 
discussion after Eqs. (\ref{69a}), (\ref{69b}).
).

Now we shall concentrate on the properties of the ground state of the pair-system 
with $n_{1,2}=0$, and 
$(r,\bar{r})=(\pm,\mp)$. 

On the above, the urciton-quark and the pair 4-momenta, respectively, 
$p_{\mu}^{(i)}$ ($i=(1,2)\rightarrow (U,\bar{V})$) 
and $P_{\mu}^{(P)}$ are related each other with {\it parton-like motion} 
$p_{\mu}^{(i)}=\kappa_{i} P_{\mu}^{(P)}$(Eq.~(\ref{EqA9a})), and given, in a simple case of single 
Flavor($\kappa_{1}=\kappa_{2}$), as:  
\begin{subequations}
\label{5-30-1}\\
~~{Momentum: }{\footnote{
Generally this equality comes from the kinematical constraint Eq.(\ref{60}). 
See, the footnote \ref{foot1}.
}}
\beq
\bs{P}^{(P)}=\sum_{i} \bs{p}^{(i)}=\bs{p}^{(1)}+\bs{p}^{(2)}
=\kappa_{1}\bs{P}^{(P)}(\bs{v})-\kappa_{2}\bs{P}^{(P)}(\bs{v})=0, 
\label{5-30-2}
\eeq

{Energy $E$:}
\beq
\hat{P}_{0}^{(P)}\equiv \hat{H}(U)+\hat{H}(\bar{V}), \ 
\hat{H}(U_{\pm})\equiv\alpha^{(1)}\bs{p}^{(1)}+\beta^{(1)}m_{\pm}^{(1)}\stackrel{(\bs{v}\to 0)}{\to}\beta^{(1)}m_{\pm}^{(1)}
 \ {\rm etc.}, \nonumber\\
 P_{0}(\bs{v})^{(P)}
 =\left\{ \begin{array}{l}
\stackrel{(\bs{v}= 0)}{=} \beta^{(1)}m_{+/-}+\beta^{(2)}\bar{m}_{-/+}
= m_{+/-}^{(1)}+\bar{m}_{-/+}^{(2)}=0 \\
\stackrel{(\bs{v} \neq 0)}{=} \xi^{(1)}(\bs{p}(\bs{v}))-\xi^{(2)}(-\bs{p}(\bs{v}))
=0 \\
\end{array} \right.,
\label{5-30-3}
\eeq
where
\beq
\xi^{(1)}(\bs{p}^{(1)})\equiv \sqrt{\bs{p}(\bs{v})_{}^2+m_{1}^2}, \ 
\xi^{(2)}(\bs{p}^{(2)})\equiv \sqrt{(-\bs{p}(\bs{v}))^2+\bar{m}_{2}^2}. \nonumber\\
\label{5-30-4}
\eeq
\end{subequations}

Thus we see that the relevant pair-system has the 4-momentum similar to that 
 of vacuum state with {\it infinite degeneracy}, as 
 \beq
 P_{\mu}^{(P)}(\bs{P}^{(P)}, P_{0}^{(P)}(\bs{v})
 )=(\bs{0},0) \ \  {\rm for} \ \ 
 (-1 <v_{i}<+1). 
 \label{5-30-5}
 \eeq
 Here it is to be noted that the relevant pair is a composite state of {\it Pauli-} and 
 {\it Chiral-} urciton (the latter is {\it very} symbolic entity of confinement). 

\paragraph{({\it Chirality Structure of Condensing Vacuum-Pair Wave Function})}

Spontaneous Breaking (SB) of chiral symmetry due to {\it vacuum condensation} 
is a well known mechanism in hadron physics. A necessary condition for the SB 
is the corresponding  {\it vacuum}/{\it ground state} has {\it Infinite Degeneracy}. 
In the last item it has been shown that 
this infinite degeneracy occurred also in our 
relevant Ground State (GS) of {\it chirality symmetric bilocal field} 
for Urciton-pair 
($U_{r\pm}, \bar{V}_{\bar{r}\mp}$) system. 
Furthermore,  in the next item we shall 
describe the {\it Dynamical Breaking} (DB) 
{\it of chirality symmetry} in the same case of urciton-pair ($U_{r\pm}, 
\bar{V}_{\bar{r}\mp}$) system (to be called the ``Vacuum-Pair System''), where the 
breaking scale (in the case of light flavor) is the mass of $\sigma$-meson, $m_{\sigma}
\approx 600$~MeV. 

In order to show the degeneracy of GS we have 
discussed the energy-momentum of 
pair-system by applying the constituent-quark spinor equation, 
{\it while here will be 
described the chirality-structure of the spin WF of Vacuum-Pair System 
by use of Urciton Spinor equation in 
chirality representation} (\ref{Eq13b}).

Our relevant local urciton-pair meson WF $\Phi_{\alpha}^{(P)\beta}(X)$ and 
its anti-pair meson WF $\Phi_{\beta}^{(\bar{P})\alpha}(X)$ 
are given by Eq.(\ref{5-29-4}), with their respective intrinsic-spin WF's, 
$W_{\alpha}^{(P)\beta}(\bs{v}(P))$ and $W_{\beta}^{(\bar{P})\alpha}(\bs{v}(P))$ 
defined in Eq.(\ref{5-29-6}), with the appropriate combination of $(r,\bar{r})$ to the 
vacuum pair. These spin-WF's are, respectively, decomposed into the eight 
elementary-members with definite attributes $\{J^{PC\chi}\}$, following the procedure 
described by Eqs. (\ref{Eq17}) and (\ref{Eq18}). 

Here, out of them, we shall select the relevant spin-WF's 
$W_{\alpha}^{(V)\beta}=\{\Gamma_{S(i)}\}_{\alpha}^{\beta}{\rm 's}$ 
to condensing vacuum-pair mesons, the scalar-type ones with 
definite and/or non-definite chirality $\Phi_{S}^{(\chi)}$'s, as follows:

These Scalar-Meson States with respective spin WF $\{\Gamma_{S(i)}\}{\rm 's}$ 
are given by 
\begin{subequations}
\label{2014-12-25-1}
\beq
\label{2014-12-25-1a}
\Phi_{S}^{Vac,(-)}(X): \ \Gamma_{S\alpha}^{(-)\beta}\equiv
\left[ 1\right]_{\alpha}^{\beta}, 
\eeq
\beq
\label{2014-12-25-1b}
\Phi_{S}^{Vac,(+)}(X): \ \Gamma_{S\alpha}^{(+)\beta}\equiv
\left[i u_{\nu}\gamma_{\nu}\right]_{\alpha}^{\beta}
\overset{(\bs{v}=\bs{0})}{\rightarrow} 
\left[-\gamma_{4}\right]_{\alpha}^{\beta}, 
\eeq
\beq
\label{2014-12-25-1c}
\Phi_{S}^{Vac,(0)}(X): \ \Gamma_{S\alpha}^{(0)\beta}\equiv
\left[ 1+i u_{\nu}\gamma_{\nu}\right]_{\alpha}^{\beta}
\overset{(\bs{v}=\bs{0})}{\rightarrow} 
\left[1-\gamma_{4}\right]_{\alpha}^{\beta}.
\eeq
\end{subequations}
Herewith, there is a relation as 
\beq
\label{2014-12-25-2}
\Phi_{S}^{Vac,(0)}(X)\approx \frac{1}{2}\left(
\Phi_{S}^{Vac,(-)}(X)\oplus\Phi_{S}^{Vac,(+)}(X)
\right),
\eeq
which{\footnote{
Some evidence for this fact will be shown in \S \ref{sec4} 
Some Phenomenological facts 
crucial to validity of the Evolved Scheme. 
A similar fact concerning the mixed-state behavior of 
the resonance $X(3872)$ 
has been pointed out by us in Ref.{\cite{Ref0}}, Proceedings of the 
1st CST-MISC Joint-Symposium. 
}} implies that the condensing vacuum-pair meson 
$\Phi_{S}^{Vac,(0)}$ is an equal-weight mixed state{
\footnote{
See,  a text book on ``Basic Quantum Mechanics'' by S. Machida\cite{Machida}. 
}} of the two-``elementary'' 
composite-hadrons, $\Phi_{S}^{Vac,(-)}$ and $\Phi_{S}^{Vac,(+)}$, with 
the mutually-opposite chirality. 

The one of two elementary vacuum-pair mesons in Eq.(\ref{2014-12-25-2}), $\Phi_{S}^{Vac,(-)}$, 
has its attribute $J^{PC\chi}=0^{++-}$. Accordingly, it may be identified with the observed meson 
``$f_{0}(500)$ or $\sigma$'' in PDG-list. This implies that the relevant chirality-symmetry 
is broken with the order{\footnote{\label{foot32}
Here it may be worthwhile to note that mass difference 
between $N(940,{\frac{1}{2}}^{+}{\rm : Nucleon})$ and $N(1440,{\frac{1}{2}}^{+}{\rm : Roper} )$, 
belonging{\cite{Ishida2002}} to the same(GS)-multiplets in the $\widetilde{U}(4)_{DS;\mathfrak{m}}$
-scheme, {\it be just of this order}.
}}
of mass, $\Delta M\approx 500 {\rm MeV}$, due to the effect of vacuum condensation. 

\paragraph{({\it Dynamical Origin of ``Broken-Chirality'' and Strong QCD-interaction})}

The origin of symmetry breaking described in the last item is due to formation of the 
mixed-state of equally-weighted two {\it elementary} composite-mesons with 
mutually opposite chirality. This is so quite-new mechanism of, might be called as, 
``Gemisch-Broken'' Chirality Symmetry, that, in the following, 
we shall sketch out its possible 
dynamical-origin in the relevant 
$\widetilde{U}(4)_{DS,\mathfrak{m}}$-classification scheme: \\
 
\noindent  i) First it is pointed out that the constituent-quark 
field $\Phi_{\alpha}(X)$ (See, Eq.(\ref{Eq10})) plays a role of the 
{\it basic asymptotic-state} of S-matrix, representing {\it those of 
general composite hadrons}: as their WF's are {\it on mass shell}, where exists the 
 ``Parton- and Parent-'' mass relation as $m^{r/\bar{s}}=\kappa^{r/\bar{s}}M$ 
 ($0<\kappa^{r/\bar{s}}<1$, $\sum_{r,\bar{s}} \kappa^{r/\bar{s}}\equiv 1$ (See, Eq.(\ref{EqA9}).) 
 
\noindent ii) Secondly it is also noted that a 
simple effective QCD-interaction between quarks is 
given by 
 \begin{subequations}
 \beq
 {\mathcal L}_{QCD}^{(I)}(\psi_{\alpha})\approx 
 \int d^4 X (\bar{\psi}(X)^{\alpha}\psi(X)_{\alpha})(\bar{\psi}(X)^{\beta}\psi(X)_{\beta}), 
 \label{2015-2-3-2a}
 \eeq
 which is conserving the chirality, although the single factor of scalar-current has 
 negative-chirality. 
 Then this $ {\mathcal L}_{QCD}^{(I)}(\psi_{\alpha})$ induces the strong interaction among 
 composite hadrons via the basic strong interaction of urciton-quarks, 
 simulator of confined quarks (inside of composite hadrons), as 
 \beq
 {\mathcal L}_{\rm str.}^{(I)}(\Phi_{\alpha})\approx 
 \int d^4 X (\bar{\Phi}(X)^{\alpha}\Phi (X)_{\alpha})(\bar{\Phi}(X)^{\beta}\Phi (X)_{\beta}). 
 \label{2015-2-3-2b}
\eeq

\noindent iii) Thirdly each pair of urciton-quarks $ \Phi_{\alpha}^{Vac,\beta}(X)\equiv \Phi_{\alpha}(X)\bar{\Phi}^{\beta}(X)$ is 
replaced by covariant 
quark-representation(, see Eq.(\ref{Eq17})), as 
\beq
\Phi_{\alpha}^{Vac, \beta}(X)=\sum_{i}\Phi^{(i)}(X)\Gamma_{\alpha}^{(i)\beta}.
\label{2015-2-3-2c}
\eeq
\label{2015-2-3-2}
\end{subequations}

\noindent iv) Then Eqs.(\ref{2015-2-3-2b}) and (\ref{2015-2-3-2c}), identifying $\Phi^{(i)}(X)$ with 
$\Phi^{Vac,(0)}(X)$, lead to the condensed vacuum $|O^{(c)}\ra$, as 
is seen in the following formulas; \\
\underline{Condensed vacuum}\\
\begin{subequations}
\label{2015-2-3-3}
\beq
\label{2015-2-3-3a}
|O^{(c)}\ra=|O_{1}^{(c)}\ra\oplus|O_{2}^{(c)}\ra, 
\eeq
\beq
\label{2015-2-3-3b}
|O_{1}^{(c)}\ra\equiv |0 \ra,  ~~|O_{2}^{(c)}\ra\equiv |\Phi_{S}^{Vac,(0)} \ra , 
\eeq
\end{subequations}
where Eq.(\ref{2015-2-3-3b}) should be understood as 
\beq
|\Phi_{S}^{Vac,(0)} \ra \approx \frac{1}{2}|\sigma^{\{+,+,-\}}\ra\oplus \frac{1}{2}|\sigma^{\{+,-,+\}}\ra.   
\eeq

\noindent v) The expectation value of chirality by condensed vacuum is easily seen to be {\it zero}, 
by noting the equation as 
\beq
\la\Phi_{S}^{Vac,(0)}|\hat{\chi}|\Phi_{S}^{Vac,(0)}\ra=
\frac{1}{2}\la \sigma|\hat{\chi}|\sigma\ra+
\frac{1}{2}\la \sigma^{'}|\hat{\chi}|\sigma^{'}\ra=\frac{1}{2}(-1)+\frac{1}{2}(+1)=0. 
\eeq
Finally in this item it should be noted that the complete set to span the relevant 
Fock-Space, in the condensed vacuum, is given by pure-vacuum, and definite $\chi$-state, 
as shown in 
\beq
\hat{1}=|0\ra\la0|+
|\sigma^{\{+,+,-\}}\ra\la \sigma^{\{+,+,-\}}|+
|\sigma^{'\{+,-,+\}}\ra\la \sigma^{'\{+,-,+\}}|. 
\eeq
\subsection{Evolved Scheme of Spectroscopy for Elementary Hadrons based on $\tau$-Quantum Mechanics} 
\label{sec3.2}
The contents, thus far presented in section \ref{sec1} and \ref{sec2}, may be 
summarized as follows: First in section \ref{sec1} the kinematical framework of 
composite hadrons is shown, {\it semi-phenomenologically}, to be 
$\widetilde{U}(4)_{DS,\mathfrak{m}}^{(q)}\otimes\widetilde{U}(4)_{DS,\mathfrak{m}}^{(\bar{q})}
\otimes O(2)_{\bs{r}\perp \bs{v}}$(see, Eq.(\ref{Eq6})); while the {\it elementary} composite hadron 
should have its new attribute $\{\chi_{m}\}$, definite chirality(see, Eq.(\ref{Eq18c})), in addition 
to the conventional $\{J,P,C\}$, as $J^{PC\chi_{m}}$. Then in section \ref{sec2} the 
$\tau$-Quantum Mechanics has been developed on a multi-quark system in 
Particle Frame with constant boost-velocity $\bs{v}$; where constituent quarks have their 
space-time coordinate $x_{\mu}^{(i)}(\tau(\bs{v}))$'s and 
the Pauli-type spin $\bs{\sigma}^{(i)}(\bs{v})$'s, 
depending on only $\bs{v}$. 
Starting from application of the variation method to an action-integral on $\tau(\bs{v})$ for 
this system, we have performed all similar to conventional procedures to 
develop Classical Mechanics, Quantum Mechanics; and the Quantum Field Theory, 
which actually leads to the relevant composite-hadron spectroscopy. 
As the result it is found that the level scheme of {\it elementary} hadrons, 
thus obtained, has the same structure as that described in section \ref{sec1}. 
Accordingly, this framework of level scheme now be called as ``Evolved Scheme of 
Spectroscopy for Elementary Hadrons''. \\

({\it Physics-Standpoint and Basic Framework of the Elementary-Hadron 
Spectroscopy} )

\noindent i) {\it Sole proper-time}(Eq.(\ref{2015-2-3-1})) through all 
constituent-quarks 
has been requisite for application of the variation method on multi-particle system. 
Furthermore, in order to get rid of the undesirable connection between the external- and 
internal- coordinates(Eq.(\ref{Eq1c})) is set the Lorentz-invariant restriction Eqs.(\ref{Eq5a}, 
\ref{Eq5b}), depending upon the boost velocity $\bs{v}$. This has made possible 
the {\it unified 1st/2nd-stage quantizations} on the internal/external coordinates of the same 
system; which is contrary to the conventional 1st/2nd-quantizations, performed 
through mutually different theoretical schemes. 

\noindent ii) {\it Particle Frame as an Inertial frame representing the situation of 
Quark-Confinement}. 
All constituents, within space-like extension, in a certain PF (with boost velocity $\bs{v}$) 
do keep their respective 4-momenta $p_{\mu}^{(r/\bar{s})}(\bs{v})\equiv \kappa^{(r/\bar{s})}
P_{\mu}(\bs{v})$ (Eqs.(\ref{2.28}) and (\ref{EqA9})), likely making free-motion. 
This is simply coming from the {\it independency} of relevant PF($\bs{v}$) from all the other 
PF($\bs{v}^{'}\neq\bs{v}$), where the corresponding 4-momenta $p_{\mu}^{'}(\bs{v}^{'})$ 
being different from $p_{\mu}^{}(\bs{v}^{})$. 

On the other hand the $\tau$-Schr{\"o}dinger Equation (\ref{2.36}) includes the solution(
see, Eqs.(\ref{215a}) and (\ref{215b})) of 
Chiralon-quark($J^{P}=\frac{1}{2}^{-}$) as a shadow particle of the 
conventional Paulon-quark($J^{P}=\frac{1}{2}^{+}$), of which two states consist the 
basic vectors of $SU(2)_{\mathfrak{m}}$-space. This freedom is the origin of new attribute 
$\{\chi\}$ of composite hadrons.

\noindent iii)  {\it Consideration within the Scope of Whole PF} with all possible 
$\bs{v}$($-1<v_{i}<1$): has shown that the multi-local $\tau$-WF's in OF is equal to the 
corresponding multi-quark WF's in $\widetilde{U}(4)_{DS,\mathfrak{m}}\otimes O(2)_{\bs{r}\perp\bs{v}}$
-scheme, which is now to be called the Evolved COQM. 
The equality Eq.(\ref{Multi-2.16b}) is reflecting that all PF's (with mutually different boost velocity 
$\bs{v}^{'}\neq \bs{v}^{''}$) have the {\it equality} as {\it Inertial Frame}, respectively. \\

({\it Concrete Framework of the Elementary-Hadron Spectroscopy})\\
\noindent iv) {\it Lorentz-Invariant Separation into the External and Internal 
coordinates} of the constituent particles has been made(see, the caption of Fig.{\ref{fig1}}), 
as 
\begin{subequations}
\beq
\label{2-9-72a}
X_{\mu}\{\bs{X}_{\rm CM}(\bs{x}^{(i)}{\rm 's}), \tau(\bs{v})\};  \ \ 
r_{\mu}^{(i)}(\equiv x_{\mu}^{(i)}-X_{\mu})
\{\bs{r}^{(i)}=\bs{x}^{(i)}-\bs{X}_{\rm CM}, r_{0}^{(i)}=0\}, 
\eeq
where 
\beq
\label{2-9-72b}
x_{\mu}^{(i)}\{\bs{x}^{(i)}, \tau(\bs{v})\} \ {\rm and} \ \bs{X}_{\rm CM}\equiv 
\sum_{i=1}^{N}m_{i}\bs{x}^{(i)}/M_0.
\eeq
\end{subequations}
This leads to the kinematical constraint on the internal 
freedom(see, Eqs.(\ref{54b}) 
and (\ref{60}) ), 
\beq
\label{2-9-73}
\sum_{i}m_{i}\bs{r}^{(i)}=0, \ \ \sum_{i}\bs{p}^{(i)}=0, 
\eeq
which has made possible the unified and separate 1st/2nd-stage quantizations 
mentioned in i). 

Here it is to be notable that the freedom  $f$ of relevant system is determined by 
the independent spatial-coordinates ($\bs{X}_{\rm CM}$ and $\bs{r}^{(i)}{\rm 's}$) of 
$N$-constituent quarks; as $f_{\rm Tot.}=f_{\rm Int.}(\bs{r}^{(i)}=\bs{x}^{(i)}-\bs{X}_{\rm CM}) 
+f_{\rm Ext.}(\bs{X}_{\rm CM})=(3N-3)+3=3N  $, where the term ($-3$) comes 
from the kinematical constraint, Eq. (\ref{2-9-73}), and implying that all the origin of $\bs{r}^{(i)}{\rm 's}$ are 
connected at $\bs{X}_{\rm CM}$. 

v) The intrinsic spin $\bs{J}$ of the relevant parent hadron comes only from those 
of constituent quarks, $\bs{S}=\sum_{i}\bs{S}^{(i)}$, while no{\footnote{\label{foot30}
It may be notable that Particle Frame of  a particle with definite orbital motion 
has {\it non-zero acceleration}, and accordingly is not {\it an inertial frame}.
\label{foot29}}} internal orbital-angular momenta, $\bs{L}=\sum_{i}\bs{l}^{(i)}$ 
($\bs{l}^{(i)}\equiv 
\bs{r}^{(i)}\times\bs{p}^{(i)}$), appear, as 
\beq
\bs{J}=\bs{S} \ \ ({\rm but \ Not \ } \bs{J}=\bs{S} \oplus \bs{L}). 
\eeq
Especially for the urciton pair-state only appear the 
elementary hadrons with the spin $\bs{J}^{(P)}=\underline{0}$
or $\underline{1}$ , as 
the scalar or 
pseudo-scalar and vector or axial-vector states. 
\beq
\bs{J}^{(P)}=\underline{\frac{1}{2}}\otimes \underline{\frac{1}{2}}=\underline{0}
\oplus \underline{1}. 
\eeq
They consists of the eight-types of elementary-mesons with the respective attribute 
$J^{PC\chi}$, as was shown in Eqs.(\ref{Eq18}). 

vi) All the radially excited states of constituent quarks contribute 
to the mass of the 
relevant parent hadrons with the oscillator quantum ($\{N^{(i)}=2n^{(i)}\}$ in the formula 
Eq.(\ref{2015-4-7-1}) as 
\beq
M_{N}^2=\sum_{i}m^2_{N^{(i)}}, \ m^2_{N^{(i)}}=m_{i}^2+N^{(i)}\Omega^{(i)}
=m_{i}^2+2n^{(i)}\Omega^{(i)} ~ (i=1,2).
\eeq
Especially for the ground state of urciton-pair mesons, see the linear-mass formula 
Eq.(\ref{4-12-2}).   
\subsection{Chirality-Symmetric, Effective Strong Interactions}
\label{sec-2015-5-26}
As it has been pointed out on i), in the last item of 
\S \ref{Degenerate Ground State}, 
(Dynamical Origin of ``Broken-Chirality'' and Strong QCD-interaction), 
that the constituent quark field 
$\Phi_{\alpha}(X)${\footnote{Here it is to be noted that, actually, the constituent quark field exist with every urcitons of the respective oscillator modes in $O(2)_{\bs{r}\perp \bs{v}}$-space. (See, 
also the footnote 24.) 
}}, defined in Eq.(\ref{Eq10}),plays a role of the {\it basic asymptotic state} 
of $S$-matrix, representing {\it those} of general, local 
composite-hadron/anti-hadron 
fields in the $\widetilde{U}(4)_{DS,\mathfrak{m}}$-spin multiplets, 
$\Phi_{\hat{N},\alpha_{1}\cdots\alpha_{n}}^{(H)\beta_{1}\cdots\beta_{m}}(X)$ / 
$\bar{\Phi}_{\hat{N},\beta_{1}\cdots\beta_{m}}^{(H)\alpha_{1}\cdots\alpha_{n}}(X)$ 
defined in (Eqs.(\ref{5-29-1}) and (\ref{5-29-2})). Then by virtue of the covariant quark 
representation, Eqs. (\ref{2014-12-25-3}), it is able to derive{\footnote{
The contents of the last item, referred above, is a simple but important example of this task. 
}} the {\it effective strong-interaction among the relevant elementary hadrons} 
$\phi^{(i)}_{\chi_{\hat{N}}(n, m)}(X)$/$\bar{\phi}^{(i)}_{\chi_{\hat{N}}(n, m)}(X)$, the members of the above multiplets. 

Its procedure is as follows: First write down 
the possible, with the ${\mathcal L}_{\rm str.}^{(I)}(\Phi_{\alpha})$ in Eq. (\ref{2015-2-3-2b}), 
urciton-quark line diagram concerning on the spin WF's ($W^{(H/\bar{H})}(\bs{v}(P))$), 
Eq. (\ref{2015-06-16-1}), of the relevant $\widetilde{U}(4)_{DS,\mathfrak{m}}$-spin multiplets. 
This diagram reflects their composite-quark structures. 
Then, taking up the trace on urciton spinor indices, it leads to the effective interaction among 
the {\it elementary-hadrons}, \{$\phi_{\chi_{\hat{N}}(n,m)}^{(i)}(X)$ 
/ $\bar{\phi}_{\chi_{\hat{N}}(n, m)}^{(i)}(X)$'s\}
(, defined in Eqs.(\ref{2014-12-25-3})), 
in the world of Local Hadrons, $H_{\hat{N},\hat{S}}(X_{\mu})$'s with 
$\widetilde{U}(4)_{DS,\mathfrak{m}}$-Symmetry, the final-stage object in Fig.{\ref{fig1}}. 
Herewith it is to be noted that {\it these effective interactions play, essentially a role of 
the S-matrix, expanded by the elementary-hadron states}, as was mentioned at the 
beginning of this subsection.

\subsection{CPT-Theorem applied to Elementary Hadrons}
\label{sec:4-4}

The validity\cite{Nishijima} of the CPT theorem is based on the invariance 
of the theory to the group of Lorentz transformations, 
the sacred spin-statistics connection, and the {\it Locality} of the theory. 
The kinematical framework, $\widetilde{U}^{(q)}(4)_{DS,\mathfrak{m}}\otimes
\widetilde{U}^{(\bar{q})}(4)_{DS,\mathfrak{m}}$-spin scheme, of 
our hadron spectroscopy satisfies all these requirement, so the theorem is 
considered to be valid, as its-self, in our scheme; except for the following revision, 
accompanied by a new attribute $\chi_{m}$ reflecting 
the $SU(2)_{\mathfrak{m}}$-symmetry, 
of elementary composite hadrons as follows:   

Applying the theorem on the conventional elementary particle-state 
$|a\rangle$, and on the composite elementary-hadron state $|H\ra$,  
respectively, as 
\begin{subequations}
\label{Eq:2015-5-26-1}
\beq
{\rm For} \ |a\ra: \ \hat{\Theta}|a\ra=|\overline{a}\ra,  
\ \hat{\Theta}\equiv\widehat{CPT}; 
\eeq
\beq
{\rm For} \ |H\ra: \ \hat{\Theta}_{H}|H\ra=\hat{\chi}_{m}|\overline{H}\ra\equiv
|\overline{H}^{(\chi_{m})}\ra,  \ \hat{\Theta}_{H}\equiv\widehat{CPT}\hat{\chi}_{m}; 
\eeq
\end{subequations}
where $\overline{a}$ represents the anti-particle of $a$ etc., ${H}^{(\chi_{m})}$ does 
the {\it chirality-partner} of $H$, and $\hat{\chi}_{m}\equiv \rho_{1}(m)$ operates on the 
urciton spinors of elementary hadrons. 

The equations (\ref{Eq:2015-5-26-1}) 
lead to the following 
relation, concerning a pair of the transition amplitudes in the mutually 
chirality-partner channels(, whose existence is guaranteed by the 
chirality symmetry), as follows:  
\beq
\la F|\hat{T}|I\ra =\la \overline{F}^{\chi_{m}}|\hat{T}^{\Theta_{H}}|\overline{I}^{\chi_{m}}\ra 
=\la I^{\chi_{m}}|\hat{T}^{\Theta_{H}\dagger}|F^{\chi_{m}}\ra, 
\eeq
where $\hat{T}^{\Theta_{H}}\equiv \hat{\Theta_{H}}\hat{T}\hat{\Theta_{H}}^{-1}$, and 
$T^{\Theta_{H}\dagger}$ represents transposed conjugate of $\hat{T}^{\Theta_{H}}$. 

This formula implies that the transformation probabilities, 
${\rm Pr.}$ and ${\rm Pr.}^{(\chi)}$ 
between all the respective states(, which are mutually chirality partners, contained in 
the chirality flow charts) 
are equal, that 
\begin{subequations}
\beq
{\rm Pr.}(I \to F)={\rm Pr.}^{({\chi})}(I^{({\chi}_{m})}\to F^{({\chi}_{m})}), 
\label{2015-06-16-2}
\eeq
where is used the equation
\beq
{\rm Pr.}\equiv|\la F|\hat{T}|I\ra|^2=\la F|\hat{T}|I\ra\la I |\hat{T}^{\dagger}|F\ra.
\eeq
\end{subequations}
The equations (\ref{2015-06-16-2}), to be called as ``Equi-Probability Relation'' 
between the transitions of chirality 
partners, and will play a some role, in the second item of sub-section 
\ref{subsec4.2}. 

\vspace{2em}

\section{Some Phenomenological Facts Crucial to Validity of the 
Evolved Hadron Spectroscopy } 
\label{sec4}

In the preceding subsection \ref{sec3.2}, 
have been summarized the essential points of the relevant 
{\it elementary-hadron spectroscopy.} Out of them in this section, the item, 
{\it concrete framework of composite-hadron spectroscopy}, will be further 
examined,{\footnote{\label{foot36}
In  this section we shall concentrate on the case of vector mesons 
of single-flavored urciton-pair, whose property is directly connected with the 
chirality symmetry. 
}}
by comparing it to the present or ingoing phenomenological status as follows: 
\subsection{Theoretical Expectations}
\label{subsec4.1}

\paragraph{({\it Non-Orbital but All Radial Excitations and Regge Trajectories})}

It leads to the eight-types of ``Regge Trajectories'' for the $(U_{r},\bar{V}_{\bar{r}})$ 
urciton-pair mesons, as was mentioned around Eqs.(\ref{Eq17}). 

{\it Types and Attributes $J^{PC\chi}$ of Mesons}
\beq
\label{Eq81}
J=\underline{0}: \ P_{s}^{(N/E)} \ 0^{-+-}/0^{-++} \ \ ; \ \ 
\ S_{}^{(N/E)} \ 0^{++-}/0^{+-+},\nonumber\\
J=\underline{1}: \ V_{\mu}^{(N/E)} \ 1^{--+}/1^{---} \ \ ; \ \ 
\ A_{\mu}^{(N/E)} \ 1^{+++}/1^{+--}.
\eeq
Here, it is to be noted that the structure of composite WF's for all 
the above pair-mesons  is basically different from that in the case of non-relativistic 
quark model. 

\paragraph{({\it Mass-Spectra of the Pair-Mesons})}

As was mentioned in subsection \ref{sec2.2.4}, the mass-squared is given by an 
independent sum of respective oscillators of each (presently $U$- and $V$-urcitons) 
constituents, $\hat{\mathfrak{m}}^{(U)2}$ 
and $\hat{\mathfrak{m}}^{(V)2}$ \\
(, see Eq.(\ref{5-29-5})).

Furthermore, in the relevant urciton-pair case, these two become identical, as a result of 
the kinematical constraints, as $\hat{\mathfrak{m}}^{(U)2} \approx
\hat{\mathfrak{m}}^{(V)2}$, see the footnote~\ref{foot36}. 
Then the $\hat{\mathcal{M}}^2_{(M)}$ for the pair-system is given as 
\begin{subequations}
\beq
\label{Eq82a}
\hat{\mathcal{M}}^2_{(M)}=\hat{\mathfrak{m}}^{(U)2}+ \hat{\mathfrak{m}}^{(V)2}=
(m_{q}^2+m^2_{\bar{q}})+(\hat{N}_{q}+\hat{N}_{\bar{q}}){\Omega}, \ \  (
\Omega^{(\bar{q})} = \Omega^{(q)} \equiv \Omega),
\eeq
which is effectively one-oscillator on one-dimensional space-vector $\bs{r}_{(M)}(\equiv 
\bs{r}^{(1)}=-\bs{r}^{(2)})$. 
\beq
\label{Eq82b}
\hat{\mathcal{M}}^2_{(M)}=m_{q}^2+m^2_{\bar{q}}+2\hat{N}_{}\Omega_{} \ 
(\hat{N}_{q}=\hat{N}_{\bar{q}}\equiv \hat{N}_{}),  \ 
{\rm and} \ {{M}}^2 (n)=m_{q^{}}^2+m^2_{\bar{q}^{}}+2{N}\Omega_{}
~(N=2n).
\eeq
Herewith it is added the linear Mass formula for the ground-state of $(U_r \bar{V}_{\bar{r}})$ 
urciton-pair mesons; as 
\beq
\label{4-12-2}
M_{(M);\rm G.S.}=m_q +m_{\bar{q}}, 
\eeq
derived by use of Eq.(\ref{4-12-4}).
Then the oscillator wave function, $H_{N}(\rho)$, is 
normalized Hermite-polynomial of $\rho$ ($\rho \equiv \sqrt{ \Omega_{q}/2} r $), 
order $N$($=0, 2, 4, \cdots$), 
of which values at the origin are given as 
\beq
\label{Eq82c}
H_{N}(0)=\frac{1}{\sqrt{2^{N-1} N! \sqrt{\pi}}}\frac{(-1)^{\frac{N}{2}}}{(\frac{N}{2})!}. 
\eeq
\end{subequations}

It is seen from Eq.(\ref{Eq82c}) that all the (ground and/or excited) radial WF's have 
non-{\it zero} value at the origin, and is expected that our relevant vector mesons have 
generally the partial decay width $\Gamma_{e^{+}e^{-}}$ into the 
electron pair with a comparable magnitude, as 
\beq
\label{Eq83}
\Gamma_{e^{+}e^{-}} \ {\rm of \ our \ relevant \ vectors}\neq 0. 
\eeq
Actually, phenomenologically, PDG\cite{PDG} reports 
that it is ``seen'' for all the vector meson with mass 
$m_{V} < 2{\rm GeV}$. Accordingly the expectation (\ref{Eq83}) seems 
to be consistent with experiments. 

\paragraph{({\it Pair-Production of Chirality-Partner Vector Mesons via Energetic-Photon $\gamma^*$})}

Firstly it is to be noted that the $\gamma^{*}$ is ignorant on the notion of 
composite-hadron chirality, while has odd ${C}$-parity. 
On the other hand 
the elementary vector-meson states $V^{(N)}$ and $V^{(E)}$ have, the mutually 
opposite, definite chirality $+$ and $-$, although they both have odd ${C}$-parity. 

Therefore, the physical situation, concerning our relevant expectation, may be 
represented through the formula, as 
\beq 
|\gamma^{*\{--, (0)\}}\rangle \to |V_{(\gamma^{*})}^{\{--,(0)\}}\rangle
\approx  \left(\frac{1}{2}|V_{(N)}^{--,(+)}\rangle\oplus\frac{1}{2}|V_{(E)}^{--,(-)}
\rangle \right).
\label{Eq84}
\eeq
This implies\footnote{
Note that this Equation (\ref{Eq84}) is analogous to Eq.(\ref{2014-12-25-2}) in the 
case of vacuum-condensation. 
} that initial $\gamma^{*}$-state with chirality ``0'' produces the final 
$V_{(\gamma^{*})}$-states, which 
is to be an equal-weight ``Gemisch''(mixed state) of 
the pure states of two-{\it elementary} composite vector-mesons, 
$V_{(N)}^{\chi (+)}$ and $V_{(E)}^{\chi (-)}$, 
as is seen from the equation 
\beq
\la V_{\gamma^{*(0)}}|\hat{\chi}|V_{\gamma^{*(0)}}\ra
=\frac{1}{2}\la V_{N}^{(+)}|\hat{\chi}|V_{N}^{(+)}\ra
+\frac{1}{2}\la V_{E}^{(-)}|\hat{\chi}|V_{E}^{(-)}\ra
=\frac{1}{2}(+1)+\frac{1}{2}(-1)=0. 
\label{20170214-1} 
\eeq
For instruction, here are given some relevant formulas, as 
\begin{subequations}
\label{Eq85}
\beq
\label{Eq85a}
V^{(1)}_{(\gamma)}(q_{+}\bar{q}_{+})
=\frac{1}{\sqrt{2}}\left(V_{(N)}(q\bar{q})+ V_{(E)}(q\bar{q})\right), 
\ V^{(2)}_{(\gamma)}(q_{-}\bar{q}_{-})
=\frac{1}{\sqrt{2}}\left(V_{(N)}(q\bar{q})- V_{(E)}(q\bar{q})\right), \ \ \\
\label{Eq85b}
V_{(N)}(q_{}\bar{q})=\frac{1}{\sqrt{2}}(V^{(1)}_{(\gamma)}(q_{+}\bar{q}_{+})+
V^{(2)}_{(\gamma)}(q_{-}\bar{q}_{-})),  \ \ 
V_{(E)}(q_{}\bar{q})=\frac{1}{\sqrt{2}}(V^{(1)}_{(\gamma)}(q_{+}\bar{q}_{+})-
V^{(2)}_{(\gamma)}(q_{-}\bar{q}_{-})). \ \ 
\eeq
\end{subequations}
where $q_{+,-}$ et al. denotes the basic {\it urciton-spinor}, 
in the Bargmann-Wigner Representation, with eigen-value 
$r(+,-)$ of mass spin $\rho_{3}(\bs{\mathfrak{m}})$(see, Eq.(\ref{Eq13a})).
Herewith Eq.(\ref{Eq85a}) points out that the two $V_{(\gamma)}$-states, 
$V_{(\gamma)}^{(1)}$ and $V_{(\gamma)}^{(2)}$, have negative $C$-parity 
and direct coupling with $\gamma^{*}$, 
while Eq.(\ref{Eq85b}) does that the latters, 
$V_{(N)}$ and $V_{(E)}$, are superposed states of the formers, 
$V_{(\gamma)}^{(1)}$ and $V_{(\gamma)}^{(2)}$, with 
the mutually-opposite relative signs. Thus one of our expectation on the 
pair-meson systems is \\
expectation: 
\beq
&&{\rm Phases \ of \ the \ coupling constant, } \ G_{V^{(1)}\gamma^{*}} ~/~ G_{V^{(2)}\gamma^{*}} \ 
{\rm between}
\ V_{(\gamma)}^{(1)} ~/~ V_{(\gamma)}^{(2)} \ {\rm with} \ \gamma^{*},  \nonumber \\ 
&&{\rm respectively, \ are} \nonumber\\
&&\hspace{8em}V^{(1)}_{(\gamma)} -\gamma^{*} \ : \ 0^{\circ} \ \ {\rm and} \ \ 
 \ V_{(\gamma)}^{(2)} -\gamma^{*} \ : \ 180^{\circ}; 
\nonumber\\
&&{\rm or \ Relative \ Sign \ between} \  V_{(\gamma)}^{(1)} \ {\rm and} \ 
V_{(\gamma)}^{(2)} \ {\rm is  \ minus}. 
\label{2016-2-23-1}
\eeq
\subsection{Phenomenological Facts}
\label{subsec4.2}

\subsubsection{Low-Mass Vector States Produced via $e^+ e^-$ annihilation}
\label{subsubsec4.2.1}

Recently the experimental members of $\sigma$-group 
performed{\footnote{\label{foot38}
The preliminary results have been 
presented at ``HADRON 2015''{\cite{Komada}}.}} 
the very cautious reanalysis of 
the $e^{+}e^{-}$ annihilation data. Their essential results in relation with 
our theoretical expectation Eq.(\ref{2016-2-23-1}) are collected in Table {\ref{tab2}}. 
\begin{table}[htb]
\caption{Mass and Phase of Low Mass Vector States Obtained in the 
Reanalysis. 
Through the relevant Experimental Process the {\it vector states} 
$(V_{\gamma}^{(1)}, V_{\gamma}^{(2)})$ 
with definite ${C}$-parity have direct coupling with $\gamma^{*}$.  
They had, thus far, been called as vector mesons; while, in Evolved-Hadron 
Spectroscopy, they are {\it Gemisch} of two {\it elementary vector mesons} ($V_{(N)}$, 
$V_{(E)}$) with same-flavor pair (See, Eqs.(\ref{Eq85a}) and (\ref{Eq85b})). 
In the Table the identification of the $V_{\gamma}$'s in the 
Evolved Spectroscopy is also given. Then it makes us possible to 
compare the expected situations on the relevant Light-mass vector-states, 
with those actually obtained in the reanalysis. 
}
\begin{tabular}{lll} \hline
\multicolumn{3}{l}{[ Experimental Process ] }\\
\multicolumn{3}{l}{$e^{+} e^{-}\to \rho(\pi^{+}\pi^{-})+\pi^{0}$} \\
\multicolumn{3}{l}{(Data) by \ ; \  
SND$(\sqrt{s}=0.66 \sim 1.38 \rm GeV)$  $\oplus$
Babar$(\sqrt{s}=1.06\sim 2.01 \rm GeV)$} \\

\multicolumn{3}{l}{[ Method of Analysis ]} \\ 
\multicolumn{3}{l}{Vector Meson Dominance with 
the $V_{(\gamma)}^{}$'s
: $[\omega (782), \phi (1020)]+\{\omega{'}(M{'}), \omega{''}(M{''})\}$}\\

\multicolumn{3}{l}{[ Values of Mass and Phase ]  obtained in Reanalysis}\\
\multicolumn{3}{l}{~~~~(Mass)  
~~$\omega (782)$ \ ~~~~~~~$\phi (1020)$ \ ~~~~~~~~$\omega{'}=\omega (1250)$
\ ~~~~($\omega (1420)^{\rm d)}$) \ 
~~~~~~$\omega{''}=\omega^{}(1650)$}\\

\multicolumn{3}{l}{~~~~\underline{(Phase) ~~~=$0^{\circ}$
 \ ~~~~~~$230^{\circ}(\simeq 180^{\circ})^{\rm b)}$ \ 
{\tiny Decisively} $180^{\circ \ {\rm a)}}$ \ ~~~({\tiny Nearly Zero-contribution}) 
\ 
$0^{\circ}$ or $180^{\circ \ {\rm c)}}$}}
\\

\multicolumn{3}{l}{(In Evolved Spectroscopy)}\\

\multicolumn{3}{l}{Identification}\\
~~~~~~~~~~~~~~~&
\underline{~~~~~~ \ ~~~~~~~~~ \ \ ~~~$1{}^{3}S_{1}$~~~~~~ \ \ ~~~ \ \ ~~~~~~~}&
~~~~~~~~~~~~~~~~~~~~~~~~~~~~~\underline{~~~$2{}^{3}S_{1}$~~~}\\
\ \ &${V}^{(1)}_{\gamma}{ (n_{+}\bar{n}_{+})}
$ ~
${V}^{(1)}_{\gamma}(s_{+}\bar{s}_{+})$ ~
${V}^{(2)}_{\gamma}(n_{-}\bar{n}_{-})$& \ 
~~~~~~~~~~~~~~~~~~~~~~~~~~${V}^{(1)}_{\gamma}(n_{+}\bar{n}_{+})$ \\

\multicolumn{3}{l}{Expected Phase}\\
~~~~~~~~~~~~~~~&  \ $0^{\circ}$(input)~~~~~~~
$\to 180^{\circ}$~~~~~~~~~~
$\to 180^{\circ}$~~~~~~
&~~~~~~~~~~~~~~~~~~~~~~~~{Either}$\left\{0^{\circ}, 180^{\circ} \right\}$
\\ 
\multicolumn{3}{l}{~~~~~~~~~~~~~~~~~~~~~{due to quark charge}~~~~~~
~~{due to $\mathfrak{m}_3$-spin}~~~~~}\\
\hline
  \end{tabular}
\label{tab2}
\end{table}

Herewith first we shall give some 
explanation on the item marked as 
a), b), c) and d) in Table {\ref{tab2}}, 
respectively, as 

a) In the energy region, $0.8~{\rm GeV} \lesssim m_{3\pi} 
 \lesssim 1.0~{\rm GeV} $, between 
 the mass of $\omega (782)$ and 
 $\phi (1020)$, which themselves have rather small-width 
$\Gamma \approx 8$ MeV and $4$ MeV, respectively,  
the relevant cross-section $\sigma_{3\pi}$ has 
a deep dip, $\sigma \sim 10$ nb, which has 
appeared as a result of destructive interference between 
their contributions $\left\{
\omega(\theta_\omega=0^{\circ}), \phi(\theta_{\phi}=180^{\circ})\right\}$.
In this situation, the contribution of 
$\omega'(1250)$ (with broad width $\sim 740$~MeV) 
in the relevant energy region is small, $\sigma\sim$ a few  
nb, but plays an important role, leading us to the very good, finer fitting. 
The main reason of this is that the phase of $\omega'(1250)$ is 
$180^{\circ}$ as a chirality partner of 
$\omega$ (with $\theta_\omega =0^{\circ}$). 
This produces constructive interference between most neighboring two states $\left\{
\phi(\theta_\phi=180^{\circ}), \omega'(\theta_{\omega'}=180^{\circ})\right\}$. 
(See, Fig.{3(a)} in Ref.{\cite{Komada}}.) 
Therefore, the choice of phase ($180^{\circ}$) on $\omega'(1250)$ 
seems to be decisive.

b) In Evolved Spectroscopy concerned only on whether the 
relative phase is $0^{\circ}$ or $180^{\circ}$ between chirality partners 
with the Same Flavor. In the relevant problem is concerned the 
two-flavor pair $(n\bar{n})$ and $(s\bar{s})$. The fitted phase 
value ($230^{\circ}$) of $\phi(1020)$, deviating from $180^{\circ}$, 
might be interpreted as a result of the $U(6)_{\sigma,F}$-broken 
 chirality symmetry(, see the last item 
 in \S \ref{Degenerate Ground State}, and footnote~{\ref{foot8}}).
 
c) In fact, both cases with relative Phases ($0^{\circ}$, $180^{\circ}$) of 
$\omega(1650)$ give 
similar good $\chi^2/{\rm d.o.f}$. This implies that the contribution of 
the state $\omega^{}(1650)$ is not interfering with those from all the other 
relevant vector-states. This situation is deduced naturally from our 
assignment; that {\it only $\omega^{}(1650)$ be radially excited state, 
while all the other vector-states} belong to {\it the ground state}. 

d) The authors of Ref. \cite{Komada}$^{\ref{foot38}}$ have 
performed both the two cases of 
the Two/Three-Resonances with \{$\omega(1250)$, $\omega(1650)$\}/
\{$\omega(1250)$, $\omega(1420)$, $\omega(1650)$\}, and found the contribution of 
$\omega(1420)$ as a $V_{(\gamma)}^{}$ is Zero, or very small. 

From all the above considerations it may be concluded that the expectation 
Eq.(\ref{2016-2-23-1}) from Evolved Hadron-Spectroscopy is consistent with the 
results of phenomenological analysis.

\subsubsection{``Riddles'' Observed by Belle Experiment around 
the Region of $\Upsilon (10860: 5 {}^{3}S_1(b\bar{b}))$}
\label{subsubsec4.2.2}

A few years ago Belle collaboration\cite{24}\cite{25} has reported the interesting 
data on Bottomonium system and pointed out a serious problem 
implied by them. 

This problem is so strange as to be called ``Riddle'' from the 
conventional Non-Relativistic classification scheme; while 
it seems us to be so natural in the Evolved Spectroscopy 
that the {\it problem itself} do disappear. 

In this item the relevant physical situation will be clarified. 
Firstly we shall study the Experimental facts summarized in 
Table.{\ref{tab3}}. 
\begin{table}[htb]
   \caption{Experimental Features of New Resonances and Identification in 
Evolved Spectroscopy. 
In {\it Experimental Data and Properties 
of New Resonances}, the properties of two spin singlet 
states ($h_{b}(1P), h_{b}(2P)$) and of two $Z_{b}$'s, are 
summarized as Experimental Fact-1, -2, -3 and Problem. 
Among them Exp. F-1/Problem corresponds to the {\it Riddles} 
referred in the title of this item. 
All of \{ Exp. F's\} might be difficult to be understood in the Non-Rel. Scheme. 
In {\it Identification of New Resonances in Evolved Classification Scheme} 
new labels of elementary meson, $A^{(E)}(mS)$/$\Upsilon^{(N)}(nS)$(, see Eq.(\ref{Eq18})), 
are assigned to $h_{b}(mP)$/$\Upsilon (nS)$. Herewith $Z_{b1}[V_{b}^{(1)}\cdot \pi]$ means 
that $Z_{b1}$ is the composed state of $V_{b}^{(1)}$ and elementary 
$\pi$ meson etc.(, see 
\S \ref{Quantum Mechanics}), and $V_{b}^{(1,2)}$ is the vector-states $V_{(\gamma)}^{(1,2)}(b\bar{b})$ defined 
in Eq.(\ref{Eq85a}). 
}
\begin{tabular}{l} \hline
\underline{Experimental Data and Properties of Observed Resonances}\\
\fbox{Two $h_b$'s: $h_b (9898)$, $h_b (10259)$}\\
Exp. F-1 \ Process: $e^{+}+e^{-}\rightarrow\Upsilon (5S) \rightarrow h_{b}(mP)\pi^{+}\pi^{-}
(m=1,2) \ \ J^P=1^{+}$\\
Production $\sigma$: Comparable to 
$e^{+}+e^{-}\rightarrow\Upsilon (5S) \rightarrow\Upsilon (nS)\pi^{+}\pi^{-}
(n=1,2,3) $\\
\underline{Problem} Why no suppression due to heavy-quark spin-flip working ? \\
(Note $\Upsilon(nS)\uparrow\uparrow$ v.s. $h_b (mP)\uparrow\downarrow$)\\
\fbox{Two $Z_{b1,2}^{\pm}(J^P =1^{+})$'s}: $Z_{b1} (10610)$, $Z_{b2} (10650)$ in Five channels:
$\Upsilon (nS)\pi^{\pm}$, $h_b (mP)\pi^{\pm}$ \\
Exp. F-2 \ Production Rate :  Similar for these 5-channles\\
Exp. F-3 \ Relat. Phase between $(Z_{b1} \ {\rm and} \ Z_{b2}) = ( 0^{\circ} \ {\rm and} \ 180^{\circ})$\\
in $\Upsilon (nS)$ and $h_b (mP)$-channel, respectively.\\
Identification of New Resonances in Evolved Classif. Scheme\\
Identif: $h_{b}((1,2)P)\to A^{(E)}((1,2)S)$; $\Upsilon (nS)\to \Upsilon^{(N)} (nS)$; 
$Z_{b1,2}\to Z_{b1}[V_{b}^{(1)}\cdot \pi]; Z_{b2}[V_{b}^{(2)}\cdot \pi] $
\\ \hline
  \end{tabular}
\label{tab3}
\end{table}

According to the new assignment, out of {\it Problem} and \{Exp. F's\}, 
the Problem do disappear, since of the relevant two-kinds of {\it different} 
intrinsic-spin(, triplet and singlet, ) particles, \{$\Upsilon(nS)$, $h_{b}(mP)$\}, now 
being regarded all as the {\it same spin-triplet} elementary hadrons. 
Herewith the (Exp. F-3) concerning the {\it relative phase} between ($Z_{b1}$, $Z_{b2}$) 
are now derived as coming from the relative phase of 
their respective constituent states ($V_{b}^{(1)}$, 
$V_{b}^{(2)}$). 

Finally, the remaining (Exp. F-1) and (Exp. F-2), 
concerning production ratio of \{$h_{b}(mP)$,$\Upsilon (nS)$\} and of 
\{$Z_{b1}, Z_{b2} $\}, respectively, 
will be also explained 
due to the {\it Chirality Symmetry} 
of the elementary hadrons. 
The relevant physical situations are concretely shown in 
Fig.{\ref{fig2}}. 
\begin{figure}[htbp]
\centering
  \fbox{{\includegraphics[width=13cm]{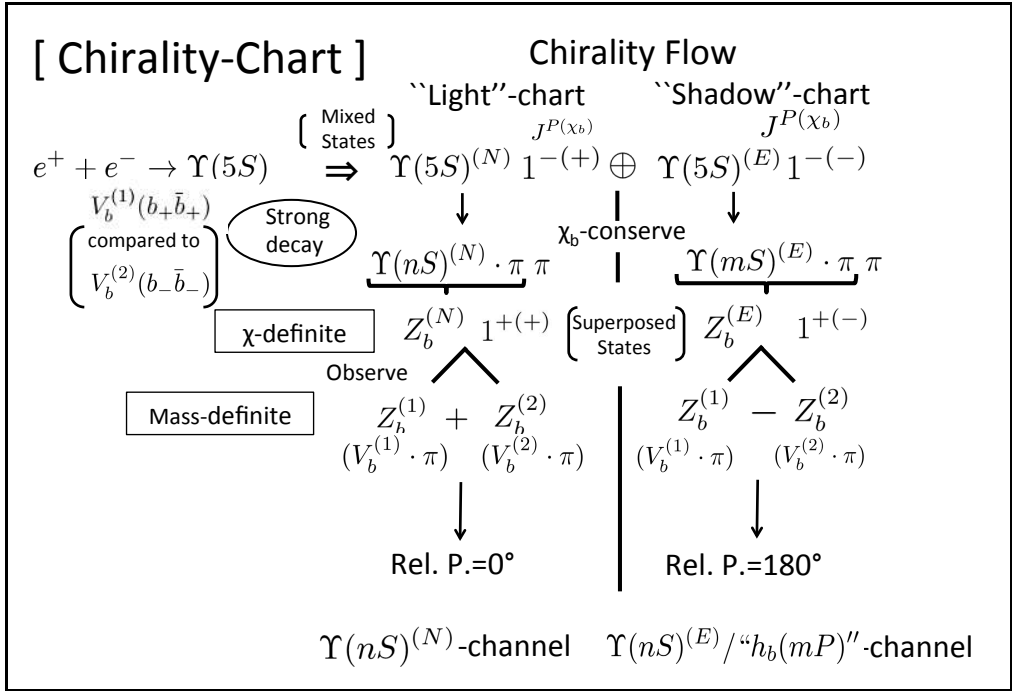}}
   }
   \caption{
Chart of Chirality-Flow through Experimental processes: 
In order to see 
concretely the relevant $\chi_{F}$-conservation is to be noticed the 
following points: 
i) Initial $\Upsilon (5S)$ state considered to be $V_{b}^{(1)}(b_{+}\bar{b}_{+};5S)$, 
is an equal-weight Gemisch of chirality-partners, $\{
\Upsilon (5S)^{(N)}, \Upsilon (5S)^{(E)}
\}$; each of which starts, respectively, the ``{\it Light}''-chart with 
$\chi_b =+$, and the ``{\it Shadow}''-chart with 
$\chi_b =-$. Then 
ii) in Light/Shadow-chart appears $\Upsilon (nS)^{(N)}\cdot \pi\pi$/
$\Upsilon (nS)^{(E)}\cdot \pi\pi$, where a new {\it elementary axial-vector} 
$J^{P(\chi_{b})}=1^{+(+/-)}$, $Z_{b}^{(N/E)}\equiv \Upsilon (nS)^{(N)}\cdot \pi
$/$\Upsilon (mS)^{(E)}\cdot \pi$, is contained. 
iii) The $Z_{b}^{(N/E)}$ is, respectively, a {\it superposed state} as $Z_{b}^{(N/E)}
=Z_{b}^{(1)}\pm Z_{b}^{(2)}$, where $Z_{b}^{(1)/(2)}$ is a {\it composite state} as 
$Z_{b}^{(1)/(2)}=(V_{b}^{(1)}\cdot \pi)/(V_{b}^{(2)}\cdot \pi)$. Herewith the 
relative phase between them (denoted as Exp. F-3) is understood as 
that of their constituent states $(V_{b}^{(1)},V_{b}^{(2)})$, as mentioned in the text. 
iv) The $\chi_{F_m}$ conservation is trivially satisfied in the relevant case, where 
always the two pions are concerned, since $\chi_{F_m}$ of $\pi$ is $\chi_{F_m}(\pi)=
-$ and  $\chi_{F_m}(\pi \cdot \pi)=(-1)^2=+$.
}
\label{fig2}
\end{figure}
Now (Exp. F-1) may naturally come from the initial state $\Upsilon (5S)$ is 
actually the equal-weight Gemisch of $\{\Upsilon (5S)^{(N)}, \Upsilon (5S)^{(E)}\}$, 
of which two are mutually chirality-partners, belonging to the same 
$\widetilde{U}(4)_{DS, 
\mathfrak{m}}$-spin multiplet. 
Then (Exp. F-2), concerning production rate for $Z_{b1}$ and $Z_{b2}$, 
may be understood as the result that the $Q$-values in their 
decay-channels, $\{\Upsilon(nS)\pi^{\pm}\}$ and 
$\{h_{b}(mP)\pi^{\pm}\}$, are almost of similar values. Herewith 
it is to be noted that, under the above interpretation of both (Exp. F-1) and 
(Exp. F-2), {\it Equi-probability Relation} between the transitions of chirality 
partners in \S \ref{sec:4-4}, has been playing a basic-theoretical role. 

\subsubsection{Another Phenomena Suggesting Non-Orbital Excitations}

In this sub-section shall be described the essential point of 
an interesting topics, which is perplexing from the conventional 
thought of line, but might be naturally understandable in the 
framework of Evolved Hadron-Spectroscopy. 

{\it Anomalous Splitting of $\psi (3770)$ in Inclusive Cross Section 
for $e^{+}e^{-}\to $ Hadrons} had been reported{\cite{26}}, 
almost a decade ago, by BES Collaboration. 
According to their cautious analysis, the line shape of cross 
sections in the energy-region between 3,700 and 3,872 GeV 
may be explained by two possible enhancements, respectively, with 
the center of mass energies of 3,764 and 3,779 GeV. Herewith also noted that 
these enhancements seem to have the relative phase 
\beq
\label{Eq86}
\phi =158\pm334\pm 5 ^{\circ}.
\eeq
{\it Interpretation from Evolved-Spectroscopy}

The above fact is understood, as that the two enhancements being the 
vector-states, $V_{(\gamma)}^{(1)}(c_{+}\bar{c}_{+})$ and 
$V_{(\gamma)}^{(2)}(c_{-}\bar{c}_{-})$, which are 
equal-weight Gemisch of elementary $(c\bar{c})$-vector mesons, 
$V^{(N)}(c_{}\bar{c}_{})$ and $V^{(E)}(c_{}\bar{c}_{})$, with definite 
$\{\chi_{m_{c},+}$ and $\chi_{m_{c},-} \}$, respectively, 
see, Eqs.(\ref{Eq84}) and 
(\ref{Eq85}). 

Then the experimental value Eq.(\ref{Eq86}) of relative phase $\phi$ is almost 
rightly to be the one, Eq.(\ref{2016-2-23-1}), expected in Evolved Hadron-Spectroscopy. 
Herewith it is also notable that in the relevant phenomena is concerned with 
the radially (possibly 1st-)excited S-wave states, and expected non-zero partial-decay 
width $\Gamma_{e^{+}e^{-}}$ (as was mentioned in Eq.(\ref{Eq83})): This expectation 
in Evolved Hadron-Spectroscopy seems to be supported 
by experiments\cite{PDG}.
\section{Concluding Remarks and Discussions} 
\label{sec5}
The purpose of this paper is to present an attempt for establishing 
a Lorentz-Invariant Composite-Hadron (to be called Elementary-Hadron) 
Spectroscopy by developing Proper-time $\tau (\bs{v})$-Quantum 
Mechanics on Multi-Quark 
System. The contents of this paper  
may be classified into the three parts as 

Part 1/~\S II and \S III : Formulation of Elementary-Hadron Spectroscopy.

Part 2/~\S V : Comparison of Elementary-Hadron Spectroscopy with Experiments. 

Part 3/~\S IV : Chirality Symmetry in Elementary-Hadron Physics.

\subsection{Specific Remarks on Elementary-Hadron Spectroscopy}
\label{sec5.2}
\subsubsection{Two Fundamental Physics Notions }
\label{sec5.2.1}
\paragraph{({\it Principle of Special Relativity on Inertial Frame})}

Firstly, here is referred to Einstein's sentence{\footnote{
The sentence is summarized by the present authors from relevant 
parts of the book{\cite{Einstein}}. 
}} on Inertial Frame as : 

\begin{enumerate}
\item[a)] Law of Nature is valid only in Inertial Frame. 
\item[b)] Law of Nature is Invariant {\it Equally} in all Inertial Frames which 
are connected, mutually, with the Lorentz-Transformation of Space-Time Coordinates. 
\item[c)] Various Inertial Frames make, mutually, {\it Rectilinear} and 
{\it Uniform} Motions. 
\end{enumerate}

In relation with the term b), it should be remarked that our method of formulating 
Elementary Hadron Spectroscopy, the $\tau (\bs{v})$-Quantum Mechanics be already 
Lorentz-invariant at the beginning. The reason is that it concerns 
all inertial frames, in the scope of, taking all possible values of $\bs{v}$, as $(-c<v_i <c)$. 
As has been discussed in \S \ref{sec3.2},
the situation of quark-confinement and strange behaviors of constituent -quarks, such 
as making Free-Particle Motion and parton-like motion, are reasonably understood from 
{\it Independency} in addition to {\it Equality}, seen in the term b), of respective 
Inertial Frame. In relation with the term c), it implies that, in Inertial Frame, there 
should not exist the states with definite orbital-angular momentum. Herewith, in 
Elementary Hadron Spectroscopy, there exist Non-orbital but All Radial Excitations, 
as has been shown by Eqs.(\ref{2.40b}) and (\ref{U-pot}) in \S \ref{new3-2-1}. 

\paragraph{({\it Conservation of Chirality in QCD-Gauge Theory and ``Elementary-Hadron''})}

The chirality is conserved through all types of the QCD interaction/upper-ground 
strong-interaction with each flavored-quarks. Therefore, somewhat elementary entity; 
the composite hadrons, consisting of  exciton-quarks with definite chirality $\{
\chi_{m_{F}}\}$, be called as {\it Elementary Hadrons}.

\subsubsection{Separation of Internal and External Space Coordinates}
\label{sec5.2.2}

The relevant separation of coordinates leads to  
\beq
X_{\mu} \ \{ \bs{X}\equiv \sum_i m^{(i)}\bs{x}^{(i)}/M_0, \tau(\boldsymbol{v})\}; \ \ 
r_{\mu}^{(i)} \ \{ \bs{r}^{(i)},0\} .
\eeq
This makes possible 
Lorentz-Invariant Unified, 1st- and 2nd-stage Quantization 
on $\{\bs{r}^{(i)}{\rm 's}, T\}$ and $\{\bs{X}, T\}$, respectively. 
They correspond to quantization of Quantum Mechanical and Field-Theoretical ones, 
respectively.

Further application of variational method on Least-Action Principle at the Particle Frame 
has led us appearance of the {\it Shadow} quark-urciton({\it Chiralon}) with $J^{P}=
{\frac{1}{2}}^{-}$, in addition to the supposed one({\it Paulon}) with 
${\frac{1}{2}}^{+}$. 
These two are basic vectors of $SU(2)_{\mathfrak{m}}$-spin space. 
The chirality 
symmetry might be broken by condensing vacuum with the 
pair of Paulon and Chiralon. 

This appearance of chiralon has deduced a vital revision on the structure of 
WF for Elementary Hadrons from the conventional WF based on 
Non-Relativistic Quantum Mechanics
\subsection{Concluding Discussions}
We have examined, in section 5, seriously the validity of Evolved 
Hadron-Spectroscopy, by comparing its expectations with the present or 
ingoing phenomenological facts. As the result, 
is obtained a fairly-promising impression. 
The framework of evolved  Hadron-Spectroscopy be completely 
Lorentz-Invariant, as is remarked in \S \ref{sec5.2.1}. 
Therefore, it seems us that 
the phenomenological knowledge, thus far obtained on the bases of 
Non-Relativistic Quark Model, should be reexamined carefully. 

\section*{Conflicts of interest}
The authors declare that they have no conflicts of interest. 
\section*{Acknowledgment}
The authors would like to express their sincere gratitude to Professors, 
K. Takamatsu, T. Tsuru, S. Naka, S. Deguchi, T. Nihei, I. Yamauchi and 
Doctor T. Komada for their useful comments and encouragement. 
Especially we owe much to Prof. Takamatsu for his insight into 
Einstein's notion on 
{\it inertial frame}, concerning the first item in section \ref{sec5.2.1}. 

The one (S.I.) of the authors is much obliged to Prof. S. Kamefuchi for his 
critical yet helpful comments. Herewith he is also grateful to Profs., 
S. Machida, Y. Ohnuki, and K. Mori for their encouragements and 
continual interests on this work. 
\begin{appendices}
\section{Lorentz Covariance for ``Local'' Composite Hadrons in the 
$\widetilde{U}(4)_{DS,\mathfrak{m}}$-Spin Scheme}
\label{AppendixA}
\paragraph{({\it Case of Dirac Particle)}}

First for instruction we recapitulate 
the relevant formulas for an elementary case of 
Dirac particle. For the infinitesimal Lorentz transformation of space-time coordinate $X_{\mu}$ 
the wave function $\Phi(X)$ transforms by the $S(\Lambda)$ as, 
\beq
\label{EqA1}
X_{\mu}^{'}&=& \Lambda_{\mu\nu}X_{\nu}, 
\ \Lambda_{\mu\nu}\equiv \delta_{\mu\nu} + \epsilon_{\mu\nu}, \\
\Phi^{'}(X^{'})&=& S(\Lambda)\Phi(X), \ S(\Lambda)=1+\frac{i}{2}\epsilon_{\mu\nu}{\Sigma_{\mu\nu}}.
\eeq
The generators $\Sigma_{\mu\nu}$, and those for sub-groups of rotation and of boost are 
given by{\footnote{
We use the Pauli-Dirac representation of $\gamma$-matrices 
$\gamma_{\mu}^{\dagger}=\gamma_{\mu} (\gamma_{i}=\rho_{2}\otimes \sigma_{i}, 
\gamma_{4}=\rho_{3}\otimes 1_{\sigma}, 
\gamma_{5}=\gamma_{1}\gamma_{2}\gamma_{3}\gamma_{4}=-\rho_{1}\otimes 
1_{\sigma})$, $\Sigma_{\mu\nu}^{\dagger}=\Sigma_{\mu\nu}, 
\boldsymbol{J}^{\dagger}=\boldsymbol{J}$ and $\boldsymbol{K}^{\dagger}=\boldsymbol{K}$.
}}
\beq
\label{EqA2}
\Sigma_{\mu\nu}=\frac{1}{2i}(\gamma_{\mu}\gamma_{\nu}-\gamma_{\nu}\gamma_{\mu}), \\
\boldsymbol{J}_{i}=\frac{1}{2}{\epsilon_{ijk}}\Sigma_{jk}, \  \boldsymbol{K}_{i}=i\Sigma_{i4}.
\eeq  
In the case of Dirac spinors with $J=1/2$ the generators and the finite transformation 
operators are explicitly given, as 
\beq
\label{EqA3}
\boldsymbol{J}_{i}&\equiv & \frac{1}{2}\sigma_{i}\otimes \rho_{0}, \ \ \ S_{R}(\boldsymbol{\theta})=
e^{-i\boldsymbol{\theta}\cdot\boldsymbol{J}}, \\
\boldsymbol{K}_{i}&\equiv &\frac{i}{2}\sigma_{i}\otimes \rho_{1}, \ \ \ S_{B}(\boldsymbol{b})=e^{-i\boldsymbol{b}
\cdot \boldsymbol{K}}, \boldsymbol{b} \equiv 
\boldsymbol{\hat{v}} \cosh ^{-1} u_{0}=\boldsymbol{\hat{v}} \cosh ^{-1} 
\left( 1/\sqrt{1-{\bs{v}}^2} \right), \label{Sb}\ \ 
\eeq
where $\theta_{i}$ are rotation angles around the $i$-axes, $u_{\mu}(\boldsymbol{v}_{i}) $ 
is the 4-velocity(3-boost velocity) of Dirac particle, and $\rho_{i}$ and $\sigma_{i}$ are the 2 by 2 Pauli matrices, representing 
the 4 by 4 Dirac matrices as $\{ \gamma \} \equiv \{ \rho \} \otimes \{ \sigma \}$. 
It may be instructive to note that 
\beq
\label{EqA4}
u_{\mu}\{u_{i},iu_{0}\}\equiv\frac{d X_{\mu}}{d \tau}=\left(\frac{\boldsymbol{v}_{i}}{\sqrt{1-\boldsymbol{v}^2}},\frac{i}{\sqrt{1-\boldsymbol{v}^2}}\right) 
\overset{(\bs{v}=\bs{0})}{\rightarrow} (\bs{0},i), 
 \\
\label{EqA5}
\boldsymbol{v}_{i} \equiv  \frac{d X_{i}}{d T} =\boldsymbol{u}_{i}/u_{0}=\boldsymbol{P}_{i}/P_{0}.
\eeq
Herewith, see Eq.(\ref{2.28}). 

\paragraph{
({\it ``Local'' Composite Hadrons})
}

The external spin-WF of composite hadrons with the 
quark-configuration $(n,m)$ ($n/m$ being the number of quarks/anti-quarks) are 
{\it tensors in the $\widetilde{U}(4)_{DS,\mathfrak{m}}$-space} 
of Dirac spinor. Accordingly the 
generators for Lorentz-transformation are given{\footnote{
It may be instructive to note that each component of tensor representation 
is transformed with the same transformation-parameters as those of basic vectors, as
$\epsilon^{(r/\bar{s})}_{\mu\nu}\equiv \epsilon_{\mu\nu}$. 
}} as 
\beq
\boldsymbol{\Sigma}_{\mu\nu}^{(n,m)}&=&\sum_{r=1}^{n}\Sigma_{\mu\nu}^{(r)}+
\sum_{\bar{s}=1}^{m}\bar{\Sigma}_{\mu\nu}^{(\bar{s})} \ 
\ (\bar{\Sigma}\equiv -\Sigma^{T}, \Sigma^{T}:{\rm Transpose \ of } \ \Sigma), 
\label{EqA6}\\
\boldsymbol{J}_{i}&=&\sum_{r=1}^{n}\boldsymbol{J}_{i}^{(r)}+\sum_{\bar{s}=1}^{m}\boldsymbol{\bar{J}}_{i}^{(\bar{s})} \  \ (\bar{J}=-J^T),\label{EqA7}\\
\boldsymbol{K}_{i}&=&\sum_{r=1}^{n}\boldsymbol{K}_{i}^{(r)}+\sum_{\bar{s}=1}^{m}\boldsymbol{\bar{K}}_{i}^{(\bar{s})} \ \ (\bar{K}=-K^T), \label{EqA8}
\eeq
where $\bar{\sum}_{\mu\nu}$ etc. denotes the complex-conjugate operator of 
$\sum_{\mu\nu}$ and so on. Herewith to be noted that, in the 1st-stage quantization, the 
WF's of quark and of anti-quark are mutually complex-conjugate functions. 
Those expressions are derived, supposing all 
``constituent''-quarks make an identical motion to that of parent, that is 
``{\it parton-like motion}'', as
\begin{subequations}
\label{EqA9}
\beq
\label{EqA9a}
p_{\mu}^{(r/\bar{s})}&=&\kappa^{(r/\bar{s})}P_{\mu}  
\ \ (0<\kappa^{\left(r/\bar{s}\right)}<1, \ \ 
\sum_{r,\bar{s}}\kappa^{(r/\bar{s})}\equiv 1), 
\eeq
where $p_{\mu}$/$P_{\mu}$ represents the four momentum of the quarks/parent. 
This leads to the relations among rotation-angles and boost-velocities of quarks/parent, as 
\beq
\boldsymbol{\theta}(\hat{\boldsymbol{p}}^{(r/\bar{s})})=\boldsymbol{\theta}(\hat{\boldsymbol{P}}); \ \ \ 
\boldsymbol{v}^{(r/\bar{s})}( = \boldsymbol{p}^{(r,\bar{s})}_{i} / p_{0}^{(r,\bar{s})}) = 
\boldsymbol{v}( = \boldsymbol{P}/P_{0}).
\label{EqA9b}
\eeq
This motion
{\footnote{
This seems to suggest a certain exciton-picture{\cite{Hara-Goto}} for quarks regarding 
space-time extension of hadrons, and gave a motive for the paper{\cite{Ref5-2}} 
which became a root of the long 
stream for the $\widetilde{U}(4)_{DS,\mathfrak{m}}$-spin classification scheme. 
See, the footnote 2.
}} 
of constituent quarks seems quite strange from a view-point of the conventional 
composite model. 
It may be worthwhile here to note that the generators for tensor representation 
Eqs.~(\ref{EqA6}) to (\ref{EqA8}) and the parton-like motion Eq.~(\ref{EqA9}) are derived directly 
for a system of multi-quarks, where the space-time coordinate of each quark $x_{\mu}^{(r/\bar{s})}$ is 
given by center of mass coordinates as 
\beq
\label{EqA10}
x_{\mu}^{(r/\bar{s})}\equiv \kappa^{(r/\bar{s})} X_{\mu}\ \ (\kappa^{(r/\bar{s})}\equiv m^{(r/\bar{s})}/(\sum_{r,\bar{s}} m^{(r/\bar{s})}) ), 
\eeq
\end{subequations} 
likely as if respective constituents shared the coordinates of 
parent themselves. 
\end{appendices}
\nocite{*}


\end{document}